\documentclass[
aps
,prb
,amsmath,amssymb,amsfonts
,superscriptaddress
,floatfix
,longbibliography
,twocolumn,10pt
]{revtex4-2}

\usepackage[T1]{fontenc}
\usepackage[utf8]{inputenc}
\usepackage{times}
\usepackage[low-sup]{subdepth}
\usepackage{wasysym}

\usepackage[usenames, dvipsnames]{color}
\usepackage[dvipsnames,svgnames,x11names,hyperref]{xcolor}

\usepackage{graphicx}

\definecolor{linkcolor}{RGB}{6,69,173} 
\definecolor{diffcolor}{RGB}{175,31,36} 
\usepackage[
  unicode=true,
  pdfusetitle,
  bookmarks=false,
  colorlinks=true,
  linkcolor=linkcolor,
  urlcolor=linkcolor,
  citecolor=linkcolor,
  pdfencoding=auto
]{hyperref}

\usepackage[
]{physpack}

\usepackage[sf]{subfigure}
\usepackage{tabularx}

\newcommand{\subfigimg}[3][,]{%
  \setbox1=\hbox{\includegraphics[#1]{#3}}
  \leavevmode\rlap{\usebox1}
  \rlap{\hspace*{10pt}\raisebox{\dimexpr\ht1-1.1\baselineskip}{#2}}
  \phantom{\usebox1}
}

\usepackage[ruled,vlined]{algorithm2e}
\SetKwInput{KwData}{Input}
\SetKwInput{KwResult}{Output}
\SetAlCapNameFnt{\small}
\SetAlCapFnt{\small}

\newcommand{\sublabel}[1]{{\footnotesize\textsf{(#1)}}}
\newcommand{\refsublabel}[1]{(#1)}

\usepackage{setspace}

\usepackage{xspace}

\begin{document}

\title{Frustration-induced emergent Hilbert space fragmentation}

\author{Kyungmin Lee}
\affiliation{Department of Physics, Florida State University, Tallahassee, Florida 32306, USA}
\affiliation{National High Magnetic Field Laboratory, Tallahassee, Florida 32310, USA}

\author{Arijeet Pal}
\affiliation{Department of Physics and Astronomy, University College London, Gower Street, London WC1E 6BT, United Kingdom}

\author{Hitesh J. Changlani}
\affiliation{Department of Physics, Florida State University, Tallahassee, Florida 32306, USA}
\affiliation{National High Magnetic Field Laboratory, Tallahassee, Florida 32310, USA}
\date{\today}

\begin{abstract}
Although most quantum systems thermalize locally on short time scales independent of initial conditions, recent developments have shown this is not always the case.
Lattice geometry and quantum mechanics can conspire to produce constrained quantum dynamics and associated glassy behavior, a phenomenon that falls outside the rubric of the eigenstate thermalization hypothesis.
Constraints “fragment” the many-body Hilbert space due to which some states remain insulated from others and the system fails to attain thermal equilibrium.
Such fragmentation is a hallmark of geometrically frustrated magnets with low-energy ``icelike manifolds'' exhibiting a broad range of relaxation times for different initial states.
Focusing on the highly frustrated kagome lattice, we demonstrate these phenomena in the Balents-Fisher-Girvin Hamiltonian (easy-axis regime), and a three-coloring model (easy-plane regime), both with constrained Hilbert spaces.
We study their level statistics and relaxation dynamics to develop a coherent picture of fragmentation in various limits of the XXZ model on the kagome lattice.
\end{abstract}

\maketitle

\section{Introduction}

\para{}
The far-from-equilibrium dynamics of interacting systems away from zero temperature shows a variety of novel phenomena.
A large class of quantum systems thermalize:
They lose memory of initial conditions and explore all corners of the many-body Hilbert space \cite{alessio_2016_ap}.
Such behavior is within the realm of the eigenstate thermalization hypothesis (ETH) \cite{deutsch_1991_pra, srednicki_1994_pre, rigol_2008_n}.
On the other hand, many-body localization provides a framework for breaking ergodicity where a disordered interacting system retains local memory of its initial conditions
\cite{basko_2006_apny, oganesyan_2007_prb, pal_2010_prb,  nandkishore_2015_arcmp, abanin_2019_rmp}.
This has germinated ideas for circumventing quantum thermalization by fragmenting the many-body Hilbert space even in the absence of disorder,
forming \emph{quantum scars}~\cite{bernien_2017_n, scherg_tiltedFH_2020, turner_2018_np, Moudgalya_NonInt_2018, Moudgalya_2018, Smith_2017, Smith_MBL_with_disorder_2017, moudgalya2019thintorus, schecter_2019_prl, choi_2019_prl, Lin_2019, Sala_2020, Khemani_2020,  chamon_2005_prl, Rakovszky2020}.
The fragmented parts fail to connect through the Hamiltonian despite being symmetry allowed, leading to slow thermalization and glassiness~\cite{Baldwin_2017, Zhihao_2018, Feldmeier2019, Lin_2020, Michailidis_2020, Pancotti_2020, feldmeier2020anomalous}.

\para{}
In recent works, quantum scars were shown to exist in a large class of frustrated spin models~\cite{lee_2020_prb, mcclarty_2020_prb, wildeboer_2020_arxiv, kuno_2020_prb},
raising the possibility of the existence of a rich variety of nonequilibrium phenomena in several magnetic systems.
Frustration in many-body systems has a long history of producing novel phases of matter whose experimental search is still ongoing.
The dynamics of excitations of frustrated systems are shown to exhibit
glassiness~\cite{chandra_1993_jpif, nussinov_2007_prb, klich_2014, Lee_2016_pnas, Gopalakrishnan_2011, cepas_2014_prb},
fractionalization and anyonic statistics~\cite{kitaev_2003_ap,kitaev_2006_ap},
and associated spin liquidity~\cite{savary_2016_rpp, sachdev_1992_prb, zeng_1990_prb, leung_1993_prb, ran_2007_prl, yan_2011_s, iqbal_2013_prb, kumar_2016_prb, plumb_2018_np}.
These phenomena are generally understood in regimes either close to the ground state or thermal equilibrium.
In this work, we shed light on situations where the system fails to explore large sections of the Hilbert space.
The tunneling between disconnected regions in Hilbert space is either entirely absent or extremely weak, which gives rise to a wide range of timescales, a hallmark of glassiness.
This form of energy-dependent hierarchy of eigenstates in frustrated systems is unraveled with energy level statistics and by mapping out connectivity and formation of ``fragments'' in the many-body Hilbert space.

\para{}
We consider a class of frustrated Hamiltonians where conservation laws emerge at low energies from local constraints,
which we broadly refer to as ``ice rules'' throughout this paper, in analogy with spin ice systems (for a review, see \cite{bramwell_2001_science}).
An exponentially large number of classical states satisfy these rules;
quantum mechanical perturbations introduce matrix elements between these states.
Since the Hamiltonian includes only local few-body operators, not every ice state is directly connected to every other ice state.
What is less obvious is that the many-body Hilbert space of ice states neatly organizes itself into isolated fragments,
a set of interconnected states with no connections to other states.
This is a pure consequence of the effective low energy behavior of the Hamiltonian,
which in turn emerges from the frustration of the lattice.

\para{}
Our focus is on two different regimes of the XXZ model on the kagome lattice
(addressed in a variety of contexts~\cite{balents_2002_prb, kumar_2014_prb, chernyshev_2014_prl, he_2015_prl,laeuchli_2015_ae, Gotze_Richter_2015, changlani_2018_prl,changlani_2019_prb} previously from the point of view of understanding its ground state):
its Ising and XY regimes characterized by macroscopically degenerate ice manifolds at low energy.
(An exponential number of quasidegenerate singlets have also been reported in the Heisenberg regime~\cite{lecheminant_1997_prb, mila_1998_prl,yao_umrigar_elser_2020}.)
The Hamiltonian is
\begin{align}
  H_{\text{XXZ}}
    &=
      \sum_{(i,j)}
        J_{ij}^{\perp} (S_{i}^{x} S_{j}^{x} + S_{i}^{y} S_{j}^{y})
        + J_{ij}^{z} S_{i}^{z} S_{j}^{z},
\label{eq:hamiltonian-xxz}
\end{align}
where $S_{i}^{\mu}$ for $\mu=x,y,z$ are spin-1/2 operators on site $i$,
and $J_{ij}^{\perp}$, $J_{ij}^{z}$ are respectively the strengths of the XY and Ising interactions on a pair of sites $(i,j)$.
While the XXZ model harbors both ice and nonice states,
we also consider its versions projected to the ice manifold.
Among the rich diversity of possible models, we study
(1) the easy-axis first-, second-, and third-nearest-neighbor XXZ model and its projected version,
introduced by Balents, Fisher, and Girvin (BFG)~\cite{balents_2002_prb},
where a ``three-up, three-down'' icelike rule on every hexagonal motif emerges,
(2) the easy-plane nearest-neighbor XXZ Hamiltonian,
with exact three-coloring ground states at $J^z_{ij}/J^{\perp}_{ij}=-1/2$~\cite{changlani_2018_prl,changlani_2019_prb},
and (3) a projected three-coloring model~\cite{cepas_2011_prb} with exponentially many three-coloring states that satisfy the rule of ``one red, one blue, and one green'' on every triangular motif.
These limiting cases of the XXZ model provide anchor points for the general phenomenology of scars and glassy dynamics of spin-$1/2$ models on the kagome lattice.

\section{Results}

\subsection{Balents-Fisher-Girvin model}

\begin{figure}\centering
\subfigimg[height=119pt]{%
\!\!\!\sublabel{a}%
}{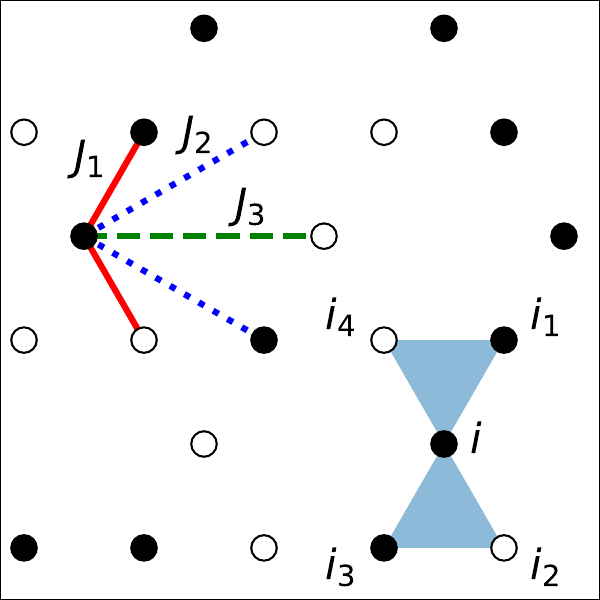}%
\quad
\subfigimg[height=113pt]{%
\raisebox{5pt}{%
\!\!\!\!\!\!\sublabel{b}%
}%
}{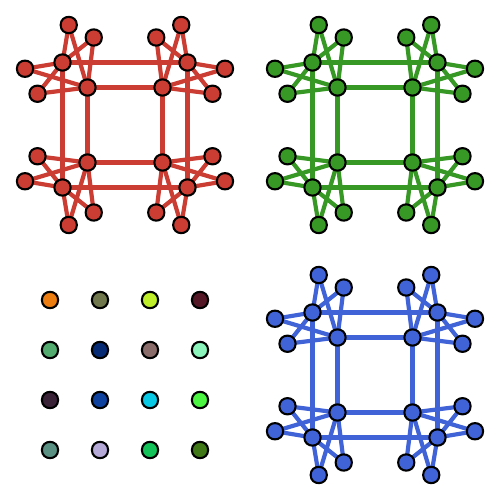}%
\\\vspace{0.5em}%
\subfigimg[width=79pt]{%
\raisebox{-5pt}{\!\!\!\!\sublabel{c}}%
}{%
{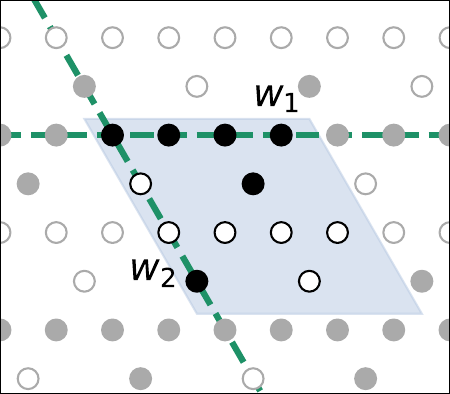}%
}
\subfigimg[width=79pt]{%
\raisebox{-5pt}{\!\!\!\!\sublabel{d}}%
}{%
{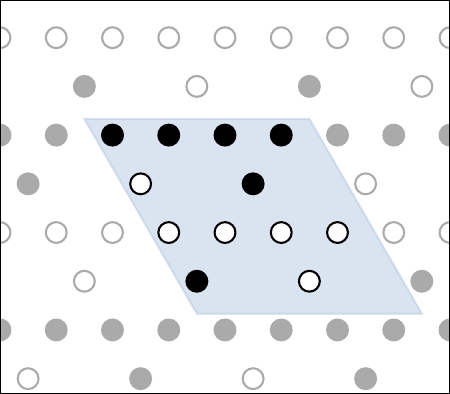}%
}
\subfigimg[width=79pt]{%
\raisebox{-5pt}{\!\!\!\!\sublabel{e}}%
}{%
{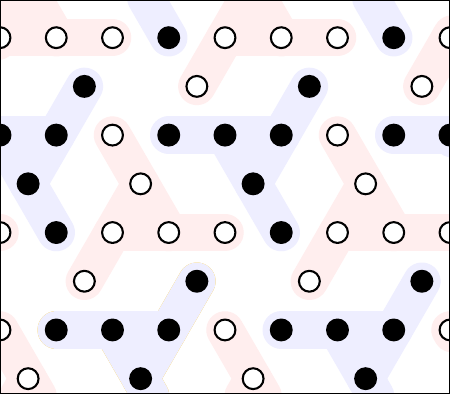}%
}
\caption{\label{fig:ice-fragmentation}%
\textbf{Structure of Balents-Fisher-Girvin Hamiltonian.}
\refsublabel{a}~%
Balents-Fisher-Girvin Hamiltonian $H_{\mathrm{BFG}}$ contains first-, second-, and third-nearest-neighbor XXZ interactions,
and the ice Hamiltonian $H_{\bowtie}$ in Eq.~\eqref{eq:hamiltonian-ice} acts on bowtie motifs.
The filled and empty circles respectively represent up and down spins in an example ice configuration (three ups and three downs for every hexagonal plaquette).
\refsublabel{b}~%
Connectivity graph of $H_{\bowtie}$ on the ice manifold of a 12-site lattice.
A vertex represents an ice configuration, a basis state for the ice manifold, and an edge represents a nonzero matrix element of the Hamiltonian between two basis states.
The three large connected components each form a topological sector,
and the 16 isolated ice configurations are the $2\mathord{\times}2$ ``triangular pinwheel'' configurations [shown in panel \sublabel{e}] related by lattice symmetry.
\refsublabel{c},~\refsublabel{d}~%
Example ice configurations on the 12-site lattice from topological sectors $(w_1, w_2) = (1,1)$ and $(1,-1)$ respectively.
$w_{i}$ are the spin parities along the two lattice directions marked by the green dashed lines in \refsublabel{c}.
\refsublabel{e}~%
$2\times2$ triangular pinwheel-ordered state.
}
\end{figure}

\begin{figure}\centering
\subfigimg[height=102pt]{%
\quad\;\;\;\sublabel{a}%
}{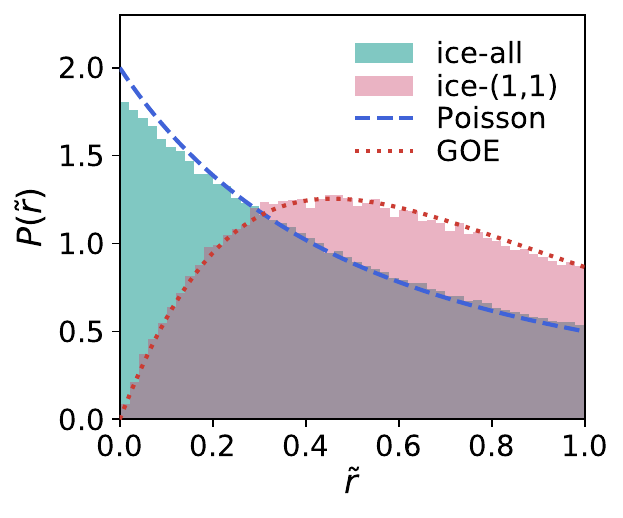}%
\;%
\subfigimg[height=102pt]{%
\quad\,\sublabel{b}%
}{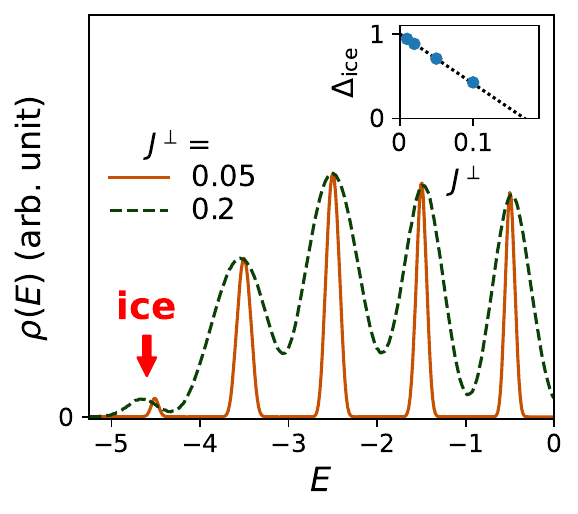}\,%
\\%
\subfigimg[height=102pt]{%
\quad\;\;\;\sublabel{c}\qquad\qquad{\footnotesize$J^{\perp}\!=\!0.01$}%
}{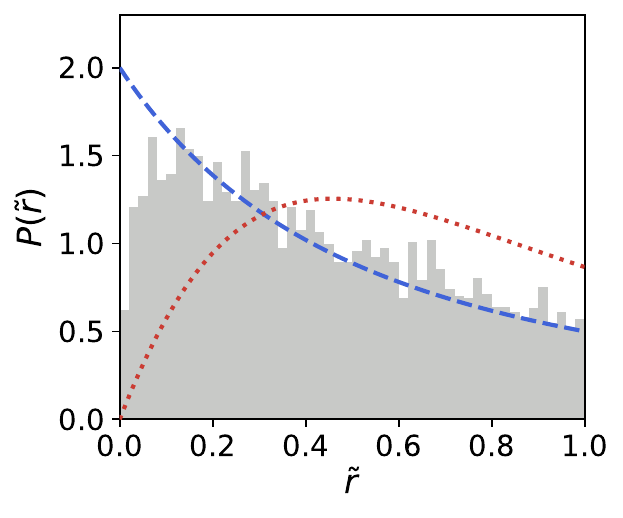}%
\!\!%
\subfigimg[height=102pt]{%
\quad\;\;\;\sublabel{d}\qquad\qquad\;{\footnotesize$J^{\perp}\!=\!0.2$}%
}{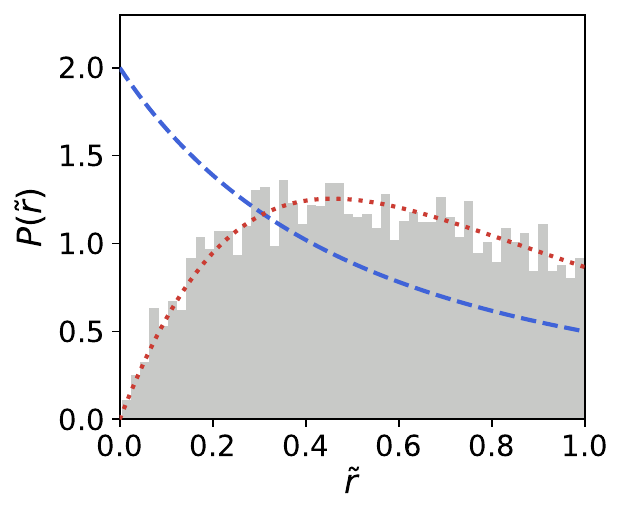}%
\caption{\label{fig:ice-level-statistics}%
\textbf{Level statistics of the BFG Hamiltonian.}
\refsublabel{a}~%
Level statistics of disordered ice Hamiltonian for a 30-site lattice,
where $J^{\bowtie}_{i}$ is site dependent, and chosen from a normal distribution centered at zero [$J_{i}^{\bowtie} \sim \calN (0, 1)$].
There are 16568 configurations in the ice manifold, partitioned into four topological sectors of the same size.
The teal (labeled ``ice-all'') and pink [labeled ``ice-(1,1)''] histograms respectively represent the probability density functions $P(\tilde{r})$ of the whole ice manifold and the topological sector $(w_1, w_2) = (1,1)$, both within the spin-flip-even sector.
\refsublabel{b}~%
Density of states of disordered BFG Hamiltonian for an 18-site lattice,
where the XY interaction is bond dependent and chosen from a uniform box distribution $J_{ij}^{\perp} \sim U(-J^{\perp}, J^{\perp})$,
while the Ising interaction is uniformly fixed to be $J_{ij}^z=1$.
The inset shows the ``ice gap''
$\Delta_{\mathrm{ice}} \equiv \min E_{N_{\mathrm{ice}}+1} - \max E_{N_{\mathrm{ice}}}$ (averaged over disorder configurations)
as a function of $J$, which closes at $J^{\perp} = J_c^{\perp} \approx 0.17$ (see Appendix~\ref{app:bfg-ice-gap}, Fig.~\ref{fig:ice-gap}).
\refsublabel{c},~\refsublabel{d}~%
Low energy level statistics of disordered BFG Hamiltonian $H_{\mathrm{BFG}}$
with nearest-neighbor XY interaction only,
for a 24-site lattice in the spin-flip-even symmetry sector.
$J^{\perp} =$ (c) 0.01, (d) 0.2.
}%
\end{figure}

\para{}
Consider the easy-axis limit $|J_{ij}^{\perp}| \ll J_{ij}^z$ of Eq.~\eqref{eq:hamiltonian-xxz}
whose sum is taken over first-, second-, and third-nearest-neighbor pairs with equal strengths, as was considered by BFG:
$H_{\mathrm{BFG}} = H_{\text{XXZ}}[J_1=J_2=J_3]$
[Fig.~\ref{fig:ice-fragmentation}(a)].
With this choice of coupling, the Ising term in the Hamiltonian becomes $H_{\text{Ising}} \propto \sum_{\hexagon} \left( \sum_{i} S^{z}_{i,\hexagon} \right)^2$
where the outer sum is over hexagonal motifs (denoted by $\hexagon$) and the inner sum is over the six sites of a given hexagon (denoted by $i,\hexagon$).
The lowest energy manifold defined by the Ising interaction consists of states with three up spins and three down spins on every hexagonal plaquette:
These states define the ice manifold of this Hamiltonian.

\para{}
The remaining XY interaction can be treated perturbatively, and contributes the following leading order term in the effective Hamiltonian of the ice manifold:
\begin{align}
  H_{\bowtie}
    &=
      \sum_{i} J_{i}^{\bowtie}
      S_{i_1}^{+} S_{i_2}^{-} S_{i_3}^{+} S_{i_4}^{-} + \mathrm{H.c.},
  \label{eq:hamiltonian-ice}
\end{align}
where the sum is over every site $i$.
$i_n$ for $n=1,2,3,4$ refer to the four sites of the bowtie motif centered at site $i$, in clockwise order [shown in Fig.~\ref{fig:ice-fragmentation}(a)].

\para{}
The Hamiltonian $H_{\bowtie}$ connects different ice configurations through tunneling.
The graph of the connection in Fig.~\ref{fig:ice-fragmentation}(b) shows multiple connected components, revealing the fragmented structure of the ice manifold.
Comparing ice configurations from different components [examples shown in Figs.~\ref{fig:ice-fragmentation}(c) and \ref{fig:ice-fragmentation}(d)],
we can identify each component as a topological sector characterized by the spin parities $w_i=\pm 1$ along the two lattice directions \cite{sheng_2005_prl}.
For large system sizes, there are four distinct topological sectors;
only three of them are allowed for the 12-site lattice due to its small size.
(We explicitly show this is the case for a larger lattice in Appendix~\ref{app:ice-appendix}.)

\para{}
In addition to the large connected components, which we associate with fragmentation, we find that there are 16 isolated states that are not part of big fragments in Fig.~\ref{fig:ice-fragmentation}(b).
These are chiral ordered states with $2\times2$ periodicity, shown in Fig.~\ref{fig:ice-fragmentation}(e), which we dub \emph{pinwheel states}.
The 16 states are related to each other through lattice symmetry and spin-flip operations.
A bowtie term in $H_{\bowtie}$ consists of two $S^+$ and two $S^-$.
Since all bowtie motifs in a pinwheel configuration have either three ups and one down, or vice versa, the state vanishes under the action of any bowtie term, and is thus an exact eigenstate of $H_{\bowtie}$ with 0 eigenvalue, which lies in the middle of the ``ice'' spectrum,
akin to quantum scars~\cite{bernien_2017_n, scherg_tiltedFH_2020, turner_2018_np, Moudgalya_NonInt_2018, Moudgalya_2018, Smith_2017, Smith_MBL_with_disorder_2017, moudgalya2019thintorus, schecter_2019_prl, choi_2019_prl, Lin_2019, Sala_2020, Khemani_2020,  chamon_2005_prl, Rakovszky2020}.
With the parent XXZ Hamiltonian in the limit $|J^{\perp}| \ll J^{z}$, these states are approximate eigenstates, and thus evolve with time.
They are, nevertheless, expected to have different dynamics than other states in the ice manifold.

\para{}
Inspired by the translationally invariant model, we study the distribution of the gap ratio
$\tilde{r}_{n} \equiv \min(s_{n}, s_{n+1}) / \max (s_{n}, s_{n+1})$~\cite{oganesyan_2007_prb},
where $s_{n}$ is the level spacing between consecutive energy levels $E_{n}$ and $E_{n+1}$,
for models with random $J_{i}^{\bowtie}$ for every site.
The randomization does not change the connectivity of Fig.~\ref{fig:ice-fragmentation}(b),
but it breaks translation and point symmetries.
This allows for easier analysis of level statistics.
We block-diagonalize $H_{\bowtie}$ with the remaining global spin-flip symmetry, which is required by the ice rule.

\para{}
Our numerical results suggest that within each topological sector, the distribution of levels follows Gaussian orthogonal ensemble (GOE) statistics, indicating chaotic dynamics within the sector
[``ice-(1,1)'' in Fig.~\ref{fig:ice-level-statistics}(a)].
Superposing the energy levels from all four distinct topological sectors results in a strong deviation from GOE, closely resembling a Poisson distribution
[``ice-all'' in Fig.~\ref{fig:ice-level-statistics}(a)].
The lack of level repulsion in the low energy manifold indicates the dramatic effect of the fragmentation of the Hilbert space, due to the ice rules emergent from frustration.

\para{}
Going back to the full BFG Hamiltonian, the conservation laws emerge at low energy for $|J^{\perp}_{ij}| \ll J^z_{ij}$, where the level statistics approaches that of $H_{\bowtie}$.
Given that the model is not exactly solvable,
it is not \textit{a priori} clear what form the level statistics acquires. To investigate how the level statistics changes at different values of $J^{\perp}$,
we study the eigenstates of disordered $H_{\mathrm{BFG}}$ with random XY interactions on the nearest-neighbor bonds whose magnitude is chosen from a uniform distribution $J_{ij}^{\perp} \sim U(-J^{\perp}, J^{\perp})$,
and $J^z_{ij}=1$ on every first-, second-, and third-nearest-neighbor bond~%
\footnote{%
Due to the small size of the lattice with 24 sites, a topological loop can be constructed with just two XY interaction terms (e.g.
one first-nearest-neighbor interaction and one second-nearest-neighbor interaction), which connects different topological sectors at order
$O((J^{\perp})^2)$ within the ``ice'' manifold.
This has the same order as $H_{\bowtie}$, and thus strongly affects the level statistics.
On the other hand, introducing XY interactions only on the first nearest-neighbor bonds, while keeping the constant $J^z$ on all first, second, and third nearest-neighbor bonds, also results in the effective Hamiltonian of the form $H_{\bowtie}$, while making topological loops $O((J^{\perp})^3)$.
For larger systems, tunneling amplitude of a topological loop becomes parametrically small, and the level statistics of $H_{\text{XXZ}}$ is expected to converge to that of $H_{\bowtie}$ for $J^{\perp} \ll J^z$, independently of whether it includes first-, second-, and third-nearest-neighbor interactions, or just the first.}.

\para{}
We find interesting level statistics in the strong Ising limit.
At a small value of $J^{\perp}$, it shows close resemblance to Poisson distribution.
This observation provides a compelling case that the fragmentation we find for the effective model survives in the full BFG model in the weak $J^\perp$ limit, given by $H_{\mathrm{BFG}}$ at low energies.
The downturn at small $\tilde{r}$ is due to topological loops which have nonzero matrix element for finite size systems (order $O((J^\perp)^3)$ here given the small size of the system), and vanish in the thermodynamic limit.
At large values of $J^\perp$, the perturbative result no longer holds as higher-order terms beyond $H_{\bowtie}$ become significant and the notion of topological sectors is no longer sharply defined.
This is indeed the case:
The low energy level statistics shows good agreement with the GOE distribution [Fig.~\ref{fig:ice-level-statistics}(d)], as the fragmented puddles are destroyed in the nonperturbative regime.

\para{}
The crossover between Poisson-like and GOE-like distributions occurs at $J^{\perp} /J_c^{\perp} \sim O(1)$, where $J_c^{\perp}$ is the value of $J^{\perp}$ at which the ice gap (defined in the caption of Fig.~\ref{fig:ice-level-statistics}) closes and the ``ice'' manifold is no longer energetically well separated from the rest of the Hilbert space.
Our finite size analysis finds no clear evidence that $J_c^{\perp}$ decreases with system size (see Appendix~\ref{app:bfg-ice-gap}).
If indeed $J_c^{\perp}$ remains nonzero in the thermodynamic limit,
Poisson-like statistics is expected for a finite range of $J^{\perp}$.

\subsection{Three-coloring model}

\begin{figure}
\begin{minipage}{74pt}
\subfigimg[width=72pt]{%
\raisebox{-2pt}{%
\!\!\!{\sublabel{a}}%
}%
}{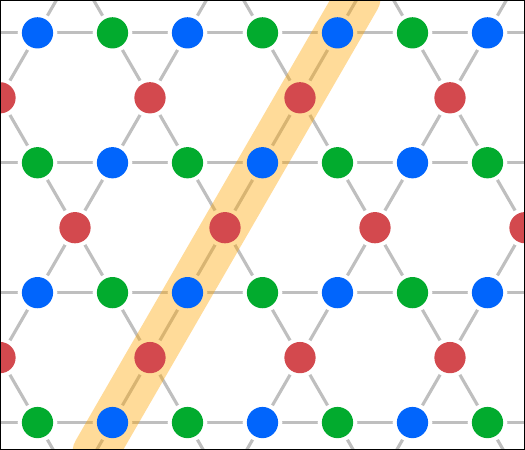}%
\\\vspace{0.5em}
\subfigimg[width=72pt]{%
\raisebox{-2pt}{%
\!\!\!{\sublabel{b}}%
}%
}{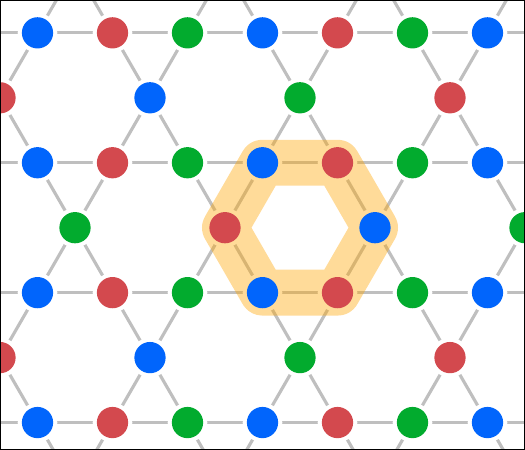}%
\\\vspace{0.5em}
\subfigimg[width=72pt]{%
\raisebox{-2pt}{%
\!\!\!{\sublabel{c}}%
}%
}{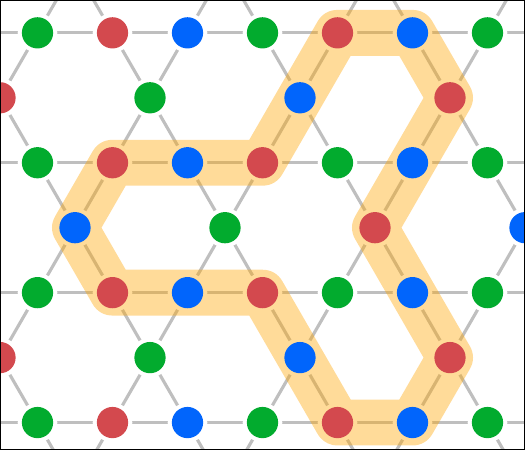}%
\end{minipage}
\begin{minipage}{168pt}\begin{flushright}%
\subfigimg[width=164pt]{%
\raisebox{-5pt}{\qquad\;\;\sublabel{d}}%
}{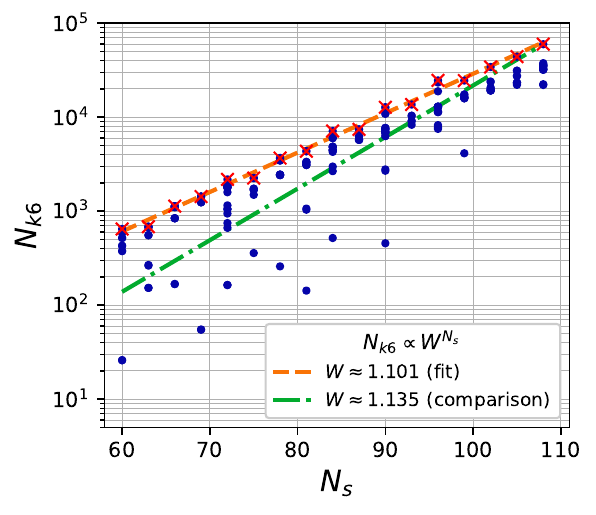}\\%
\subfigimg[width=166pt]{%
\raisebox{-2pt}{\qquad\;\;\,\sublabel{e}}%
}{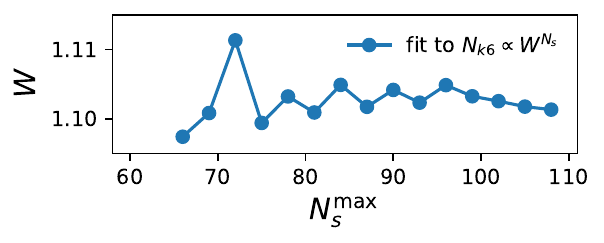}%
\end{flushright}\end{minipage}%
\caption{\label{fig:three-coloring}%
\textbf{Three-coloring states and Kempe loops.}
\refsublabel{a},~\refsublabel{b},~\refsublabel{c} Three-coloring states of a kagome lattice.
\refsublabel{a} and \refsublabel{b} are respectively referred to as $q\mathord{=}0$ and  $\sqrt{3}\mathord{\times}\sqrt{3}$ states.
The orange lines example red-blue Kempe loops:
In the $q\mathord{=}0$ state, Kempe loops are all \emph{global}, while in the $\sqrt{3}\mathord{\times}\sqrt{3}$ state, they are all \emph{local}.
\refsublabel{d}
Number of connected components in Kempe connectivity graph with loops of length 6 ($N_{k6}$) vs number of sites ($N_s$), for various system geometries.
The orange dashed line is an exponential fit $N_{k6} \sim W^{N_{s}}$ to the largest number of components for every system size (marked by a red cross).
For comparison, we present an exponential with the $W$ for the scaling of the number of colorings \cite{baxter_1970_jmp} (shown as the green dashed-dotted line), normalized to match the data point at 108 sites.
\refsublabel{e}
Scaling analyses for the exponential fits.
The $y$ axis represents the parameter $W$ from fitting to data points in the range $60 \le N_{s} \le N_{s}^{\mathrm{max}}$.
}
\end{figure}

\begin{figure*}\centering
\begin{minipage}{245pt}
\subfigimg[width=240pt]{%
\raisebox{6pt}{%
\!\!\!\!\!\!\!\sublabel{a}%
}%
}{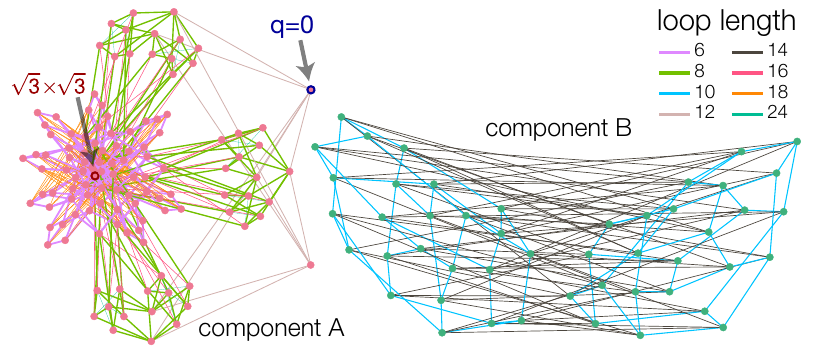}
\subfigimg[width=240pt]{%
\raisebox{12pt}{%
\!\!\!\!\!\!\!\sublabel{b}%
}%
}{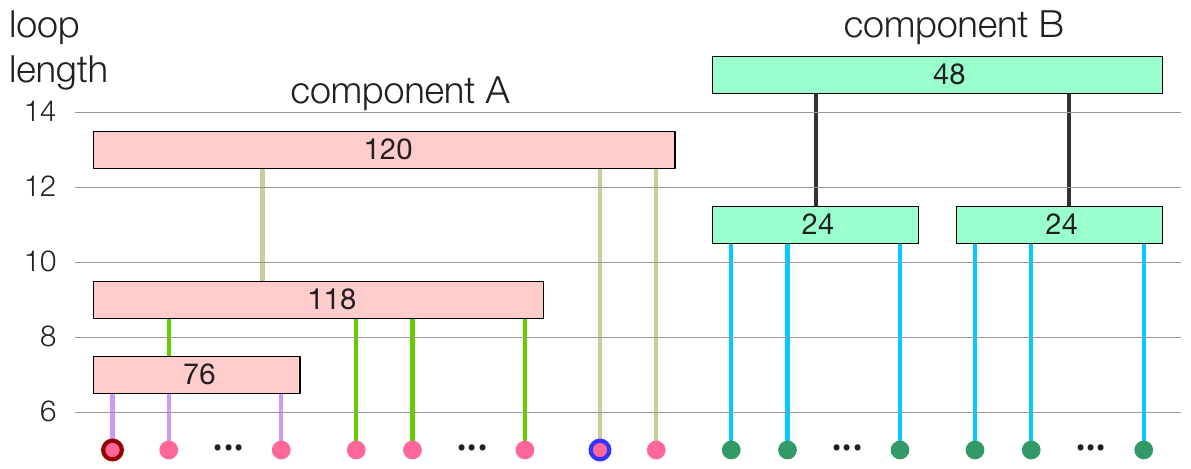}%
\end{minipage}\quad
\begin{minipage}{245pt}
\subfigimg[width=240pt,height=225pt]{%
\raisebox{-4pt}{\!\!\!\sublabel{c}}%
}{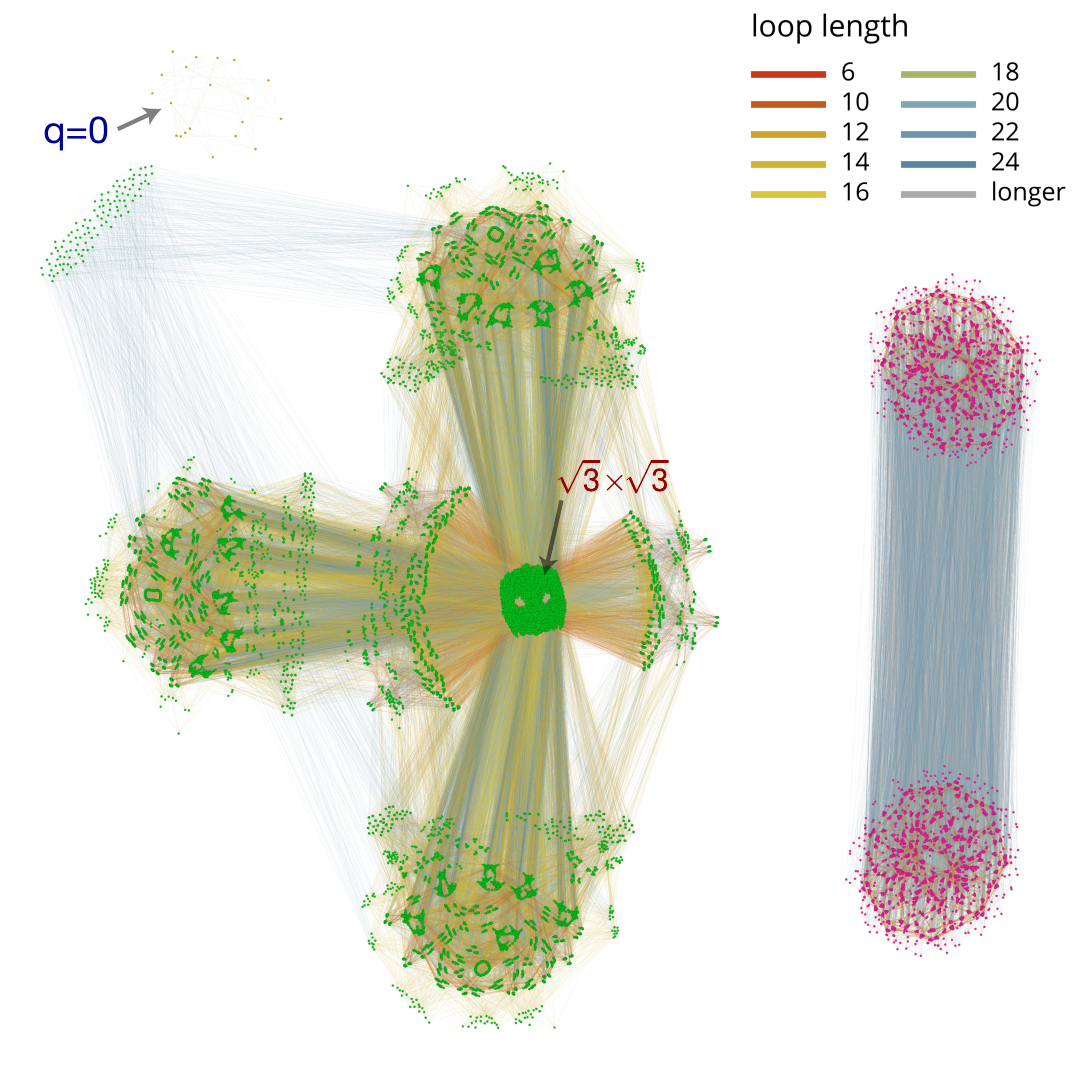}%
\end{minipage}
\caption{\label{fig:kempe-connectivity}%
\textbf{Kempe connectivity of three-coloring manifold.}
\refsublabel{a}
Kempe connectivity graph of the three-coloring manifold of a 36-site lattice (168 colorings).
A vertex represents a three-coloring, and an edge represents a Kempe move that connects two colorings.
The color of a vertex represents the component it belongs to, and the color of an edge represents the length of the move.
Two special configurations are highlighted by blue and red circles,
$q\mathord{=}0$ and $\sqrt{3}\mathord{\times}\sqrt{3}$, respectively shown in Figs.~\ref{fig:loschmidt}(a) and \ref{fig:loschmidt}(b).
The graph clusters into two connected components (labeled A and B) when loops of all lengths are allowed.
\refsublabel{b}
The hierarchical component structure of the graph in \refsublabel{a}.
A leaf node (circle) represents a three-coloring,
and an internal node (rectangle) represents a connected component when Kempe moves of a certain length or shorter (represented by the gray horizontal lines) are allowed.
Vertical lines mark component-subcomponent relationship;
a child node is a subcomponent of its parent node.
The number shown in an internal node is the number of colorings in the connected component.
The colors of vertices and edges match those of panel \refsublabel{a}.
\refsublabel{c}
Kempe connectivity graph of the three-coloring manifold on a 81-site lattice (45,184 colorings).
Length dependent structures are presented in the Supplemental Material~\cite{sm} Video 1.
}
\end{figure*}

\begin{figure*}%
\subfigimg[height=130pt]{%
\raisebox{0pt}{%
\qquad\;\sublabel{a}}%
}{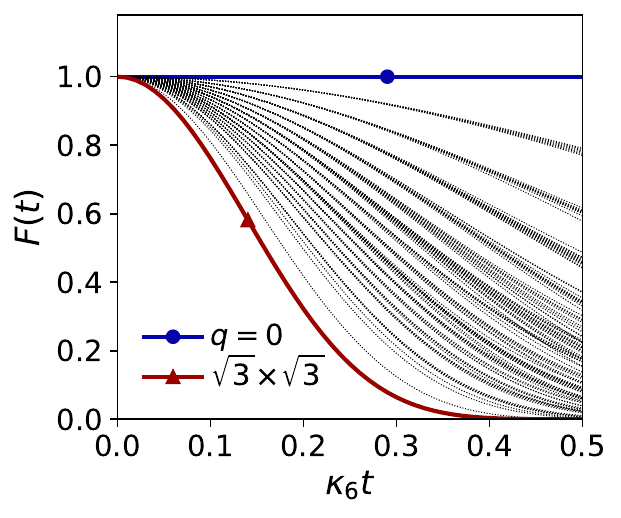}%
\;\;%
\subfigimg[height=130pt]{%
\raisebox{0pt}{%
\qquad\;\;\sublabel{b}}%
}{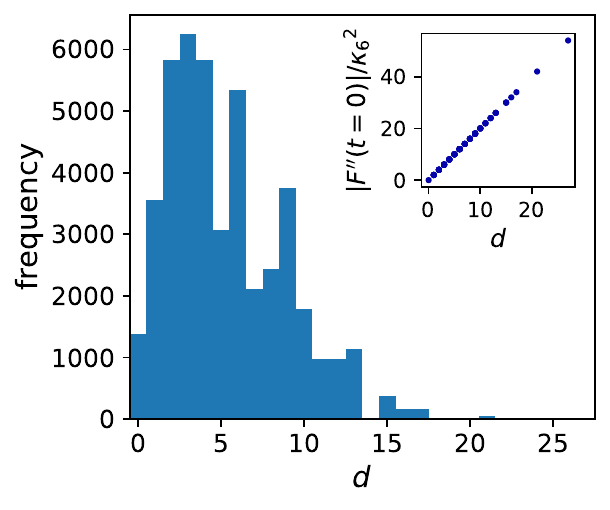}
\quad%
\subfigimg[height=130pt]{%
\raisebox{0pt}{%
\qquad\;\sublabel{c}}%
}{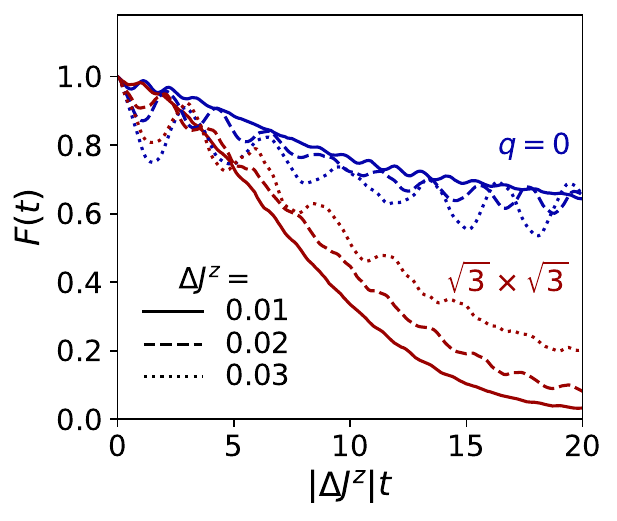}
\caption{\label{fig:loschmidt}
\textbf{Dynamics of three-coloring states.}
\refsublabel{a}
Loschmidt echoes
$F(t) \equiv \vert \langle \psi(0) \vert \psi(t) \rangle \vert^2$
for length-6 Kempe model on a 81-site kagome lattice.
The blue and red curves respectively represent Loschmidt echoes of
$q\mathord{=}0$ state and $\sqrt{3}\mathord{\times}\sqrt{3}$ state.
The black dotted curves are Loschmidt echos of other coloring states.
\refsublabel{b}
Degree histogram of the Kempe connectivity graphs with loops of length 6, i.e., distribution of numbers of Kempe loops $d$ for all three-colorings.
Inset: $F''(t)$ vs. number of Kempe loops, confirming the relationship $F(t) \approx 1 - d {\kappa_6}^2 t^2$.
\refsublabel{c}
Loschmidt echoes for XXZ model $H_{\text{XXZ}}^{\text{n.n.}}$ at $\Delta J^z \equiv J^z + 1/2 =0.01$, 0.02, 0.03, with 18 sites.
}
\end{figure*}

\para{}
The BFG description was designed to explore the Ising regime of the kagome antiferromagnet:
The low energy manifold defined by its ice rule breaks up into four large connected components, along with isolated scar states.
A complementary viewpoint is provided by a model in the XY regime of the XXZ Hamiltonian, which also contains an exponentially large low energy manifold defined by local constraint. 
Contrary to the BFG model, however, these states belong to an exponentially large number of fragments, with a hierarchical structure, as we will see in the following.

\para{}
The Hamiltonian of the second model also takes an XXZ form:
\begin{align}
  H_{\mathrm{XXZ}}^{\mathrm{n.n.}}
    &=
      \sum_{\langle i, j \rangle}
        S_{i}^{x} S_{j}^{x} + S_{i}^{y} S_{j}^{y}
      + J^{z} S_{i}^{z} S_{j}^{z}
\end{align}
where the sum is now only over the nearest neighbor pairs $\langle i,j\rangle$, and the XY coupling strength is set to $J^{\perp}=1$.
At $J^z=-1/2$, the ground states are exactly known for the kagome lattice~\cite{changlani_2018_prl}:
They are the exponentially many three-coloring states~\cite{chalker_1992_prl, sachdev_1992_prb, harris_1992_prb, huse_1992_prb, henley_2009_prb,delft_1992_prl}, each of which satisfies the constraint of exactly one red, one blue, and one green degree of freedom
[$\left(\ket{\up} + \exp(2\pi i n/3) \ket{\dn}\right) / \sqrt{2}$ with $n=0,1,2$ respectively]
on every triangular motif. The exact solutions are tensor products of these degrees of freedom, which remain exact ground states under projection to any $S_z$ sector~\cite{changlani_2018_prl, changlani_2019_prb}.

\para{}
Two representative three-colorings shown in Figs.~\ref{fig:loschmidt}(a) and \ref{fig:loschmidt}(b), the $q\mathord{=}0$ and $\sqrt{3}\mathord{\times}\sqrt{3}$ states, have periodic structures and are relevant for the ground state and finite-temperature phase diagram of the kagome antiferromagnet~\cite{changlani_2018_prl,tokuro_2016}.
Three representative three-colorings are shown in Figs.~\ref{fig:three-coloring}(a)--\ref{fig:three-coloring}(c).
The first and second states, respectively referred to as $q\mathord{=}0$ (three-site unit cell) and $\sqrt{3}\mathord{\times}\sqrt{3}$ states (nine-site magnetic unit cell), have periodic structures and are relevant for the ground state and finite-temperature phase diagram of the kagome antiferromagnet~\cite{changlani_2018_prl,tokuro_2016}.
These states are characterized by the differences in their two-color loops (\emph{Kempe loops}):
In the $q\mathord{=}0$ case, the Kempe loops wind around the torus, and in the $\sqrt{3}\mathord{\times}\sqrt{3}$ case, they are local ones of  length 6.
Exchanging the two colors within a loop also yields a valid three-coloring,
since the sites adjacent to the chain all have the third color which is different from the two colors of the loop.
Different colorings have Kempe loops of different lengths and some colorings contain both local loops and global loops [see Fig.~\ref{fig:three-coloring}(c)].
We also note that, unlike the Ising ice states, three-coloring wavefunctions are \emph{not orthogonal} to each other, but there is an intricate structure of how they are connected to one another under two-color loop moves (\emph{Kempe moves}).

\para{}
To explore the structure of the three-coloring manifold with Kempe loops, we construct a model where
  (1) the non-three-coloring states are completely projected out,
  and (2) the allowed three-colorings are assumed to be orthogonal to each other
    (akin to the status of dimer coverings in the Rokhsar-Kivelson model~\cite{rokhsar_kivelson}).
We then impart dynamics to the three-colorings by Kempe moves for loops of certain lengths.
The resulting effective model is a Hamiltonian \cite{castelnovo_2005_prb, cepas_2011_prb} in the many-body Hilbert space of three-colorings which is written as
\begin{align}
  H_{K}
    &=
      \sum_{C} \sum_{k \in K(C)}
      \kappa_{\ell_k} \vert k(C) \rangle \langle C \vert,
\end{align}
where the sum is over every three-coloring $C$.
$K(C)$ is the set of Kempe loops in $C$,
$\ell_{k}$ is the length of the Kempe loop $k$,
and $k(C)$ is the resulting three-coloring after exchanging the two colors within the Kempe loop $k$ from $C$.
(Here we define the colorings as equivalent up to global ``color rotations,'' e.g., red to green, green to blue, blue to red,
which corresponds to projecting to a particular $S_z$ sector in the XXZ model.)

\para{}
We find that, contrary to the BFG model,
the number of fragments in this model scales exponentially with system size.
In Fig.~\ref{fig:three-coloring}(d), we show number of fragments (i.e., connected components)
vs. number of sites, for Kempe connectivity graphs including loops of length 6 only, for all inequivalent lattice geometries having 60 to 108 sites.
The number of fragments vary substantially even for lattices with the same number of sites,
especially for small systems which are strongly influenced by the boundary condition.
We can nevertheless estimate the scaling $N_{k6} \sim W^{N_{\text{s}}}$ from the largest number of fragments for each system size, where $N_{k6}$ and $N_s$ are respectively the number of fragments and the number of sites.
We find that $W\approx 1.101$, which is smaller than the $W \approx 1.135$ for the scaling of the number of three-colorings \cite{baxter_1970_jmp}.
For comparison, we also tried fitting the same data points to a power law (shown in Appendix~\ref{app:kempe-connectivity} Fig.~\ref{fig:kempe-component-vs-size-power}).
Although the limited range of system sizes makes it difficult to conclusively distinguish an exponential from a power law,
finite size scaling of the fitting parameters strongly suggests that the number of fragments scales exponentially.

\para{}
Since there are Kempe loops of various lengths,
the length-dependent connectivity of the three-coloring states
shows a rich structure.
(See Appendix~\ref{app:kempe-algorithm} for algorithms used to generate the connectivity graph.)
Figure~\ref{fig:kempe-connectivity}(a) shows an illustrative example of the connectivity graph of the three-coloring states of the 36-site lattice with full point group symmetry 6mm of the infinite lattice.
When loops of all lengths are allowed, we identify two connected components:
There are no Kempe moves that connect a state from component A to a state from component B.
The fragmentation structure is much richer on larger lattices, shown in Fig.~\ref{fig:kempe-connectivity}(c) for 81 sites.
Both structures nevertheless show close resemblance,
e.g. there are tightly bound sets of states in the center to which the $\sqrt{3}\mathord{\times}\sqrt{3}$ configurations belong, and
the $q\mathrm{=}0$ configurations are weakly connected to other states at longer loop lengths.

\para{}
The length-dependent hierarchy of the Hilbert space fragments is shown in Fig.~\ref{fig:kempe-connectivity}(b) for the 36-site lattice.
This hierarchy provides a way of organizing the states in terms of slow and fast modes.
States which connect with each other at shorter loops thermalize faster, compared to those that require longer loops (i.e., higher up in the hierarchy).
In the thermodynamic limit, there should be exponentially many fragments for finite loop lengths, and hence a broad distribution of relaxation times.
This hierarchically constrained dynamics is analogous to classical glasses~\cite{palmer_1984_prl},
where the relaxation due to fast modes involving the stronger bonds is constrained by the slow modes.

\para{}
Within this framework, one would expect the $q\mathord{=}0$ state,
which has only ``topological'' Kempe loops and hence is an exact eigenstate (unless $\ell_{k}$ is macroscopically large)
to relax parametrically slower than the $\sqrt{3}\mathord{\times}\sqrt{3}$ state which has  loops of length 6.
This is confirmed by full diagonalizations of the effective model on an 81-site kagome lattice.
In Fig.~\ref{fig:loschmidt}(a), we plot the overlap between the time-evolved wavefunction and the initial state, also known as Loschmidt echo, for various coloring states, with $\kappa_{\ell}=1$ for $\ell=6$, and $0$ otherwise.
In addition to confirming our intuition for the $q\mathord{=}0$ and $\sqrt{3}\times\sqrt{3}$ states, we observe that, for other states with a mixture of Kempe loops of length 6 and longer, the relaxation time scales are in the intermediate range.
For small $t$, each curve is expected to follow $\sim \cos^2(\sqrt{d} \kappa_6 t)$, where $d$ is the number of Kempe loops, i.e., the degree of a node in the Kempe connectivity graph.
(See Appendix~\ref{app:kempe-dynamics} for proof.)
Based on this relationship, we can use $d$ as a proxy for the relaxation dynamics at short times.
The histogram in Fig.~\ref{fig:loschmidt}(b) shows that $d$ forms a broad distribution.
Further analyses on the distribution of the relaxation time scales presented in Appendix~\ref{app:kempe-dynamics}
suggest exponential number of states with glassy relaxation dynamics in the thermodynamic limit.

\para{}
To see whether our findings on the varying time scales for the coloring states applies to
the full XXZ Hamiltonian,
we consider the time evolution of the $q\mathord{=}0$ and $\sqrt{3}\mathord{\times}\sqrt{3}$ states projected to the $S_z=0$ sector,
under $H_{\mathrm{XXZ}}^{\mathrm{n.n.}}$.
(We also carried out calculations for the unprojected coloring states.
See Appendix~\ref{app:unprojected-coloring} for a discussion on the effect of $S_z$ projection.)
In Fig.~\ref{fig:loschmidt}(b), we plot the Loschmidt echo
for the two coloring states on an 18-site lattice, calculated with full diagonalization.
This is used to diagnose the relaxation times of the coloring states.
At $J^z=-1/2$, they are exact eigenstates, and hence do not evolve with time.
Away from, but close to, this special point,
these states thermalize at distinctly different rates despite being at similar energies:
The $\sqrt{3}\mathord{\times}\sqrt{3}$ state thermalizes much faster than the $q=0$ state.
For both, the time scale is approximately proportional to $\Delta J^z \equiv J^z+1/2$.
(The difference appears to get smaller as one goes even further away from $J^z=-1/2$.)
The fact that the thermalization time depends on the Kempe loop structure of the three-coloring state confirms a qualitative explanation in terms of the phenomenological model.

\section{Conclusion}

\para{}
In conclusion, we have presented a class of spin-$1/2$ models on the kagome lattice,
whose low energy quantum dynamics, governed by icelike rules, forces the formation of Hilbert space fragments, giving rise to glassy dynamics as a consequence.
This includes easy-axis (Ising) and easy-plane (XY) regimes both of which harbor macroscopically degenerate manifolds.

\para{}
Our work opens several questions related to the nonequilibrium dynamics of frustrated spin systems.
The dynamics of defects in the constrained ice manifold could potentially be useful for protecting information in excited states of many-body systems~\cite{Huse_LPQO_2013, bahri_2015_nc}.
The relationship between the graphical structure of fragmented clusters and the emergent symmetries in the model can shed light on the dynamics of symmetry breaking in the presence of frustration and can provide an alternative route to understanding the ground states in these models~\cite{balents2010SL}.
Finally, the recent developments in synthetic quantum systems of Rydberg atoms and trapped ions in two dimensions have made possible the experimental realization of these models in the different regimes.
(See Refs.~\cite{Gyu-Boong_Stamper-Kurn_2012, hazzard_2014_pra,browaeys_2020_np} for review.)
Local addressability and monitoring of the quantum states in these experiments are promising for preparing these three-coloring states and investigating their dynamics.

\begin{acknowledgments}
We thank V. Elser, E. Yuzbashyan, A. Patra, O. Vafek, V. Dobrosavljevic, K. Hazzard, and S. Pujari for valuable discussions on a wide range of related topics.
H.J.C. thanks G. Chan for encouraging him to think about nonequilibrium dynamics of frustrated magnets.
K.L. and H.J.C. acknowledge support from Florida State University and the National High Magnetic Field Laboratory.
The National High Magnetic Field Laboratory is supported by the National Science Foundation through NSF/DMR-1644779 and the state of Florida.
H.J.C. was also supported by NSF CAREER Grant No. DMR-2046570.
A.P. was funded by the European Research Council (ERC) under the European Union's Horizon 2020 research and innovation programme (Grant Agreement No. 853368).
We also thank the Research Computing Cluster (RCC) at Florida State University for computing resources.
\end{acknowledgments}

\appendix

\section{Convention for lattice geometry}

\begin{figure}[b]%
\includegraphics[width=200pt]{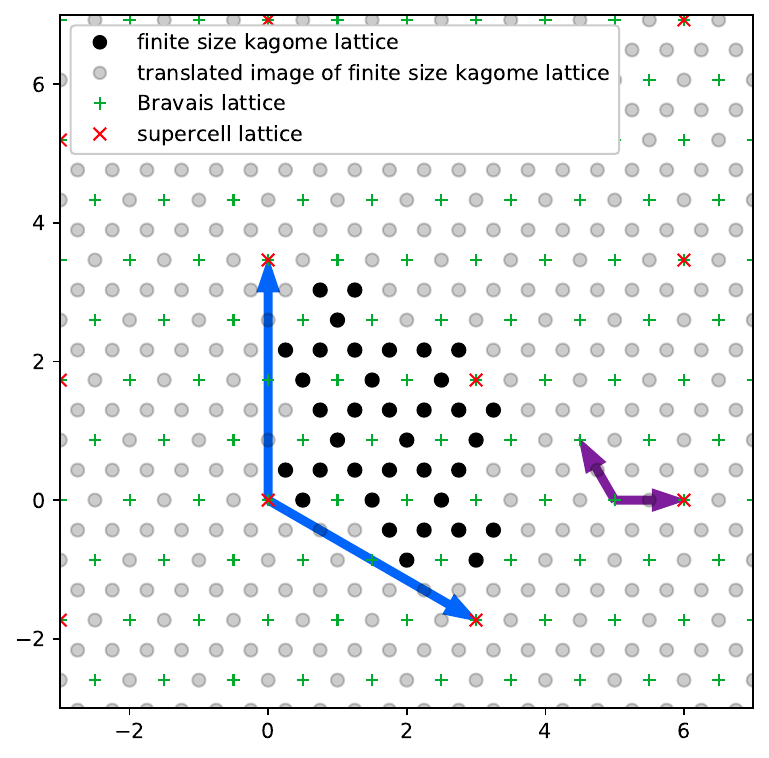}
\caption{\label{fig:geometry-36}%
Geometry of a finite size kagome lattice with shape $(2,-2)\times(2,4)$.
The purple arrows on the right mark the lattice vectors $\bfa_1$ and $\bfa_2$ of the Bravais lattice (represented by green pluses).
The blue arrows are the lattice vectors of the superlattice (represented by red crosses):
$\bfb_1 = 2 \bfa_1 - 2\bfa_2$ and $\bfb_2 = 2\bfa_1 + 4\bfa_2$.%
}%
\end{figure}

\para{}
Throughout the paper, we study kagome lattices of various sizes.
Here we briefly explain the convention used in this work for specifying the geometry of a finite size lattice.

\para{}
A finite size system with periodic boundary condition can be thought of as a quotient space of an infinite space by superlattice translations.
Therefore, one way to specify the geometry of the finite size lattice is by defining the superlattice through its lattice vectors.

\para{}
A kagome lattice is defined by a triangular Bravais lattice with a basis of three sites.
Throughout this work we use Bravais lattice vectors $\bfa_1 = (1,0)$ and $\bfa_2 = (-1/2, \sqrt{3}/2)$.
The superlattice vectors can be written in units of these lattice vectors.
For example, $(2,-2)\times(2,4)$ refers to a 36-site cluster, shown in Fig.~\ref{fig:geometry-36}.
$(2,-2)$ and $(2,4)$ respectively represent the superlattice vectors $\bfb_1 = 2\bfa_1 - 2\bfa_2 = (3,-\sqrt{3})$ and $\bfb_2 = 2\bfa_1 + 4\bfa_2 = (0, 2\sqrt{3})$.

\para{}
Shapes of the kagome lattices used in the paper are listed in Table~\ref{tab:geometry}.
\begin{table}[b]\centering%
\caption{\label{tab:geometry}Geometries of the kagome lattices used in the paper.}
\begin{ruledtabular}
\begin{tabular}{ccc}
Lattice size & Shape & Used in\\
\hline
12 & $(2,0)\times(0,2)$ & Fig.~1(b)--1(d)\\
18 & $(3,0)\times(1,2)$ & Figs.~2(b), 5(c)\\
24 & $(2,-1)\times(2,3)$ & Figs.~2(c), 2(d)\\
30 & $(3,-1)\times(1,3)$ & Fig.~2(a)\\
36 & $(2,-2)\times(2,4)$ & Figs.~4(a), 4(b)\\
81 & $(3,-3)\times(3,6)$ & Figs.~4(c), 5(a), 5(b)
\end{tabular}
\end{ruledtabular}
\end{table}

\section{Balents-Fisher-Girvin ``ice'' model}
\label{app:ice-appendix}

\begin{figure}
\begin{minipage}{190pt}%
\subfigimg[width=185pt]{\raisebox{0pt}{\sublabel{a}}}{{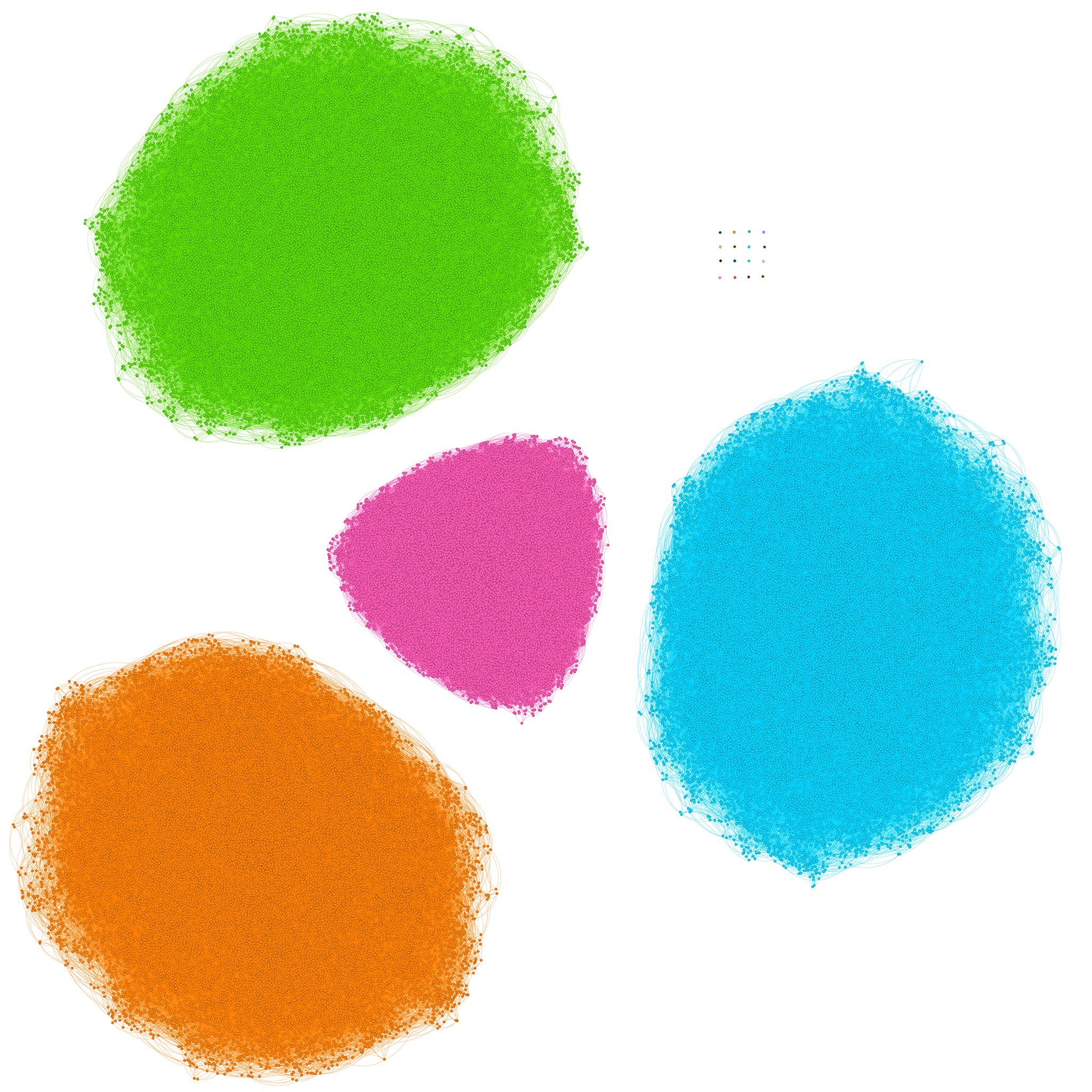}}
\end{minipage}
\vspace{1em}
\begin{minipage}{244pt}%
\subfigimg[width=115pt,page=1]{\raisebox{-7pt}{\!\sublabel{b}}}{{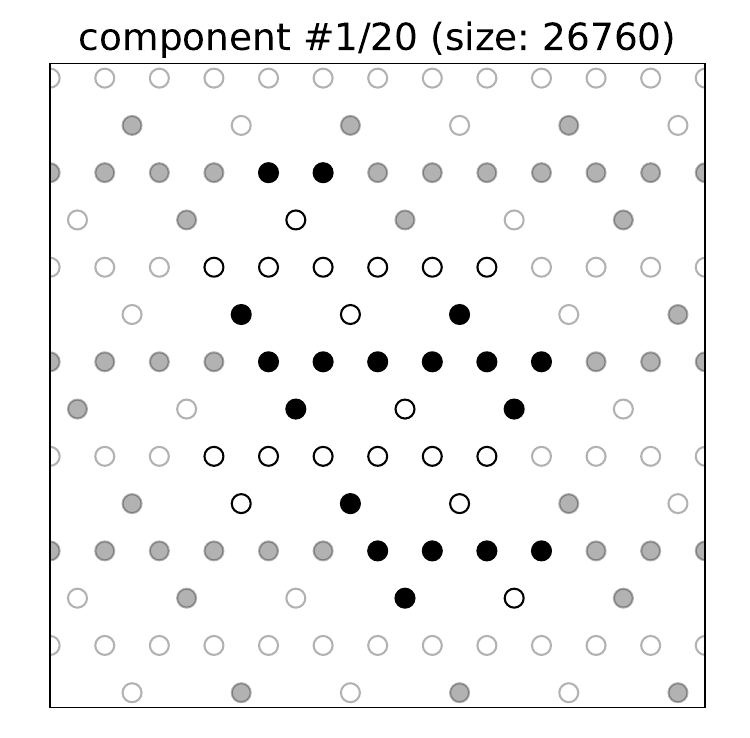}}%
\subfigimg[width=115pt,page=2]{\raisebox{-7pt}{\!\sublabel{c}}}{{ice-configurations_36.pdf}}
\subfigimg[width=115pt,page=3]{\raisebox{-7pt}{\!\sublabel{d}}}{{ice-configurations_36.pdf}}%
\subfigimg[width=115pt,page=4]{\raisebox{-7pt}{\!\sublabel{e}}}{{ice-configurations_36.pdf}}%
\end{minipage}%
\caption{\label{fig:ice-config}%
\refsublabel{a}
Connectivity graph of $H_{\bowtie}$ on Balents-Fisher-Girvin ice manifold on a 36-site kagome lattice with 107,176 ice configurations.
Each vertex represents an ice configuration, and an edge represents a non-zero tunneling element between two ice configurations that the edge connects.
The ice manifold fragments into four topological sectors, shown as four connected components in the graph.
The color of a vertex indicates the connected component it belongs to.
\refsublabel{b}-\refsublabel{d}
Representative configurations in each of the four topological sectors (chosen arbitrarily),
for $(2,-2)\times(2,4)$ cluster.
Opaque black circles (open circles for down spins, and filled circles for up spins) mark the sites within the cluster, and gray circles represent translated image sites under periodic boundary condition.
Corresponding values of topological invariant $(w_1, w_2)$ are
\refsublabel{b} $(1,-1)$,
\refsublabel{c} $(1,1)$,
\refsublabel{d} $(-1,1)$, and
\refsublabel{e} $(-1,-1)$.
}
\end{figure}

\para{}
The ice manifold of Ref.~\cite{balents_2002_prb} has four topological sectors characterized by topological invariants $w_i=\pm 1$, the spin parities along two lattice directions.
Such breaking up of the Hilbert space can be visualized as multiple connected components in the connectivity graph of the Hamiltonian $H_{\bowtie}$ on the ice manifold.
In the main text we have shown the graph for a 12-site cluster, which showed only three connected components, in addition to 16 isolated states.
This, however, was due to the small size of the system.
Figure~\ref{fig:ice-config}(a) is the connectivity graph for a 36-site system
with 107,176 ice configurations, which clearly shows \emph{four} distinct connected components, in addition to the 16 isolated states.
Representative ice configurations (arbitrarily chosen) of the four topological sectors are shown in
Figs.~\ref{fig:ice-config}(b)--\ref{fig:ice-config}(e).

\begin{figure*}
\subfigimg[width=100pt]{\raisebox{0pt}{\quad\sublabel{a}}}{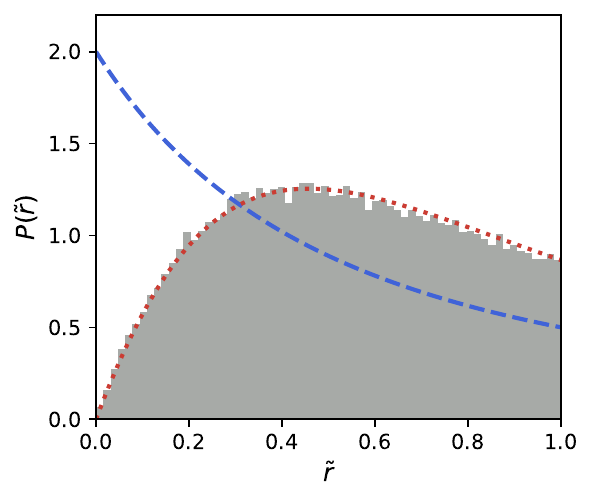}%
\subfigimg[width=100pt]{\raisebox{0pt}{\quad\sublabel{b}}}{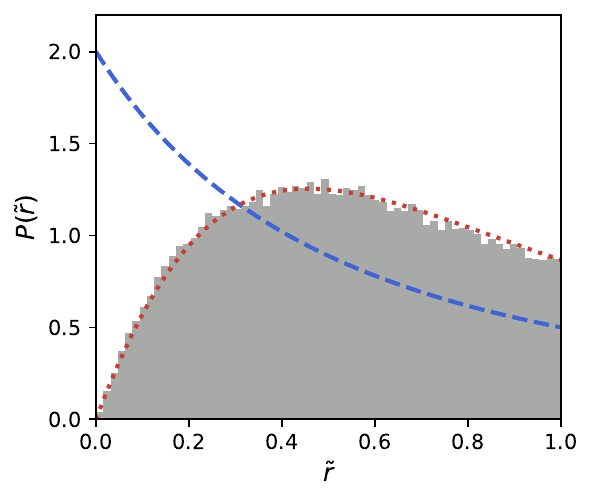}%
\subfigimg[width=100pt]{\raisebox{0pt}{\quad\sublabel{c}}}{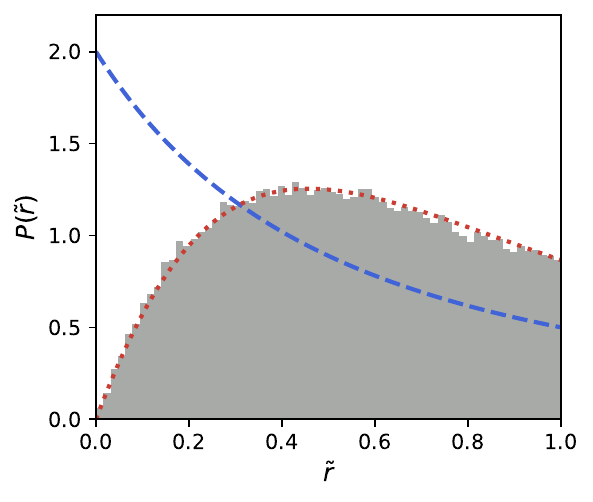}%
\subfigimg[width=100pt]{\raisebox{0pt}{\quad\sublabel{d}}}{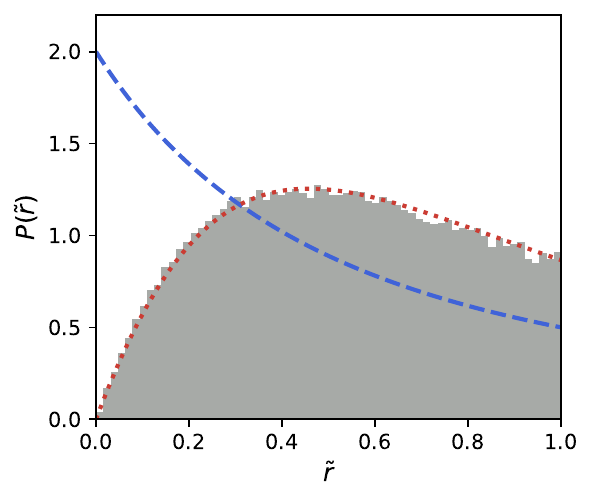}\\%
\subfigimg[width=100pt]{\raisebox{0pt}{\quad\sublabel{e}}}{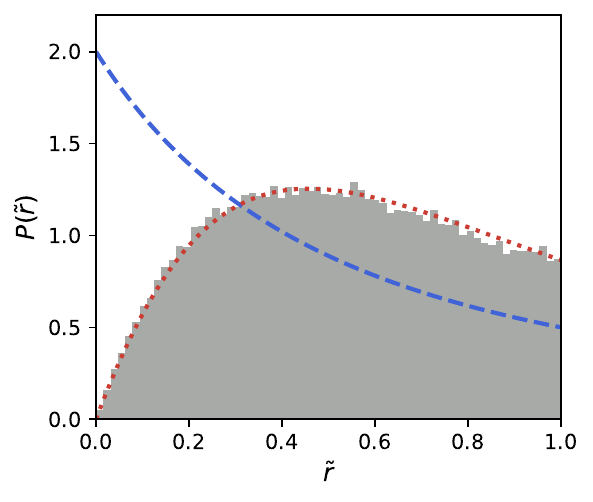}%
\subfigimg[width=100pt]{\raisebox{0pt}{\quad\sublabel{f}}}{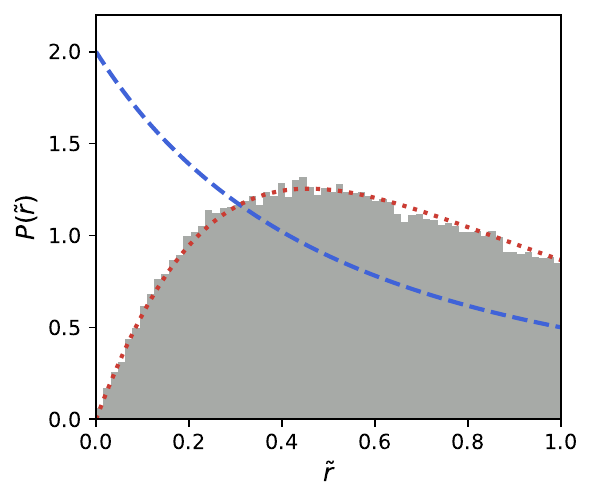}%
\subfigimg[width=100pt]{\raisebox{0pt}{\quad\sublabel{g}}}{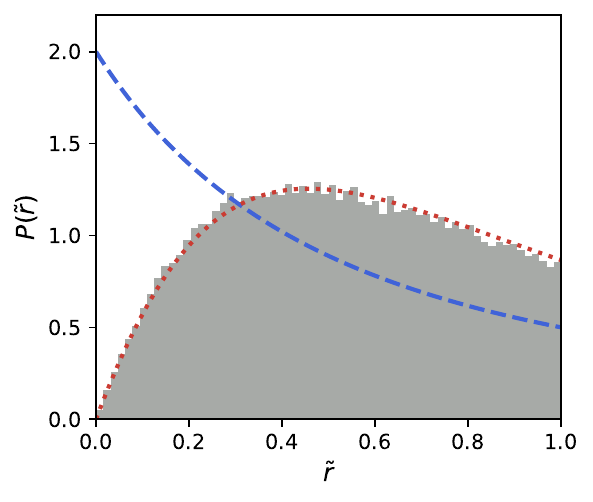}%
\subfigimg[width=100pt]{\raisebox{0pt}{\quad\sublabel{h}}}{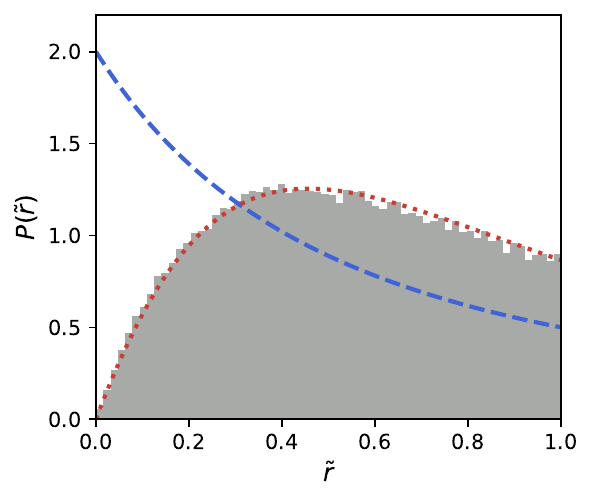}\\
\subfigimg[width=130pt]{\raisebox{0pt}{\quad\;\;\sublabel{i}}}{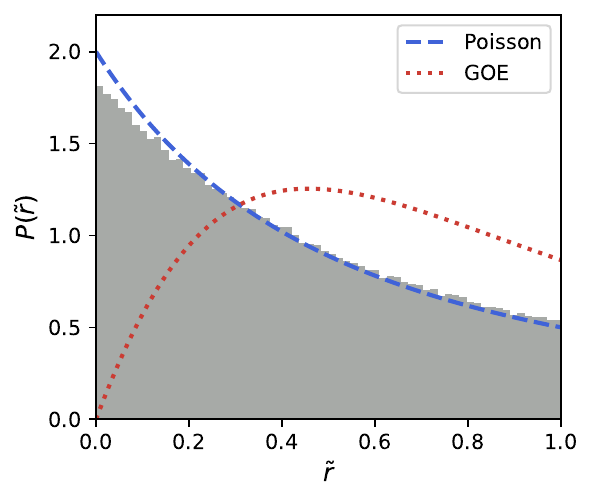}\;\;%
\subfigimg[width=130pt]{\raisebox{0pt}{\quad\;\;\sublabel{j}}}{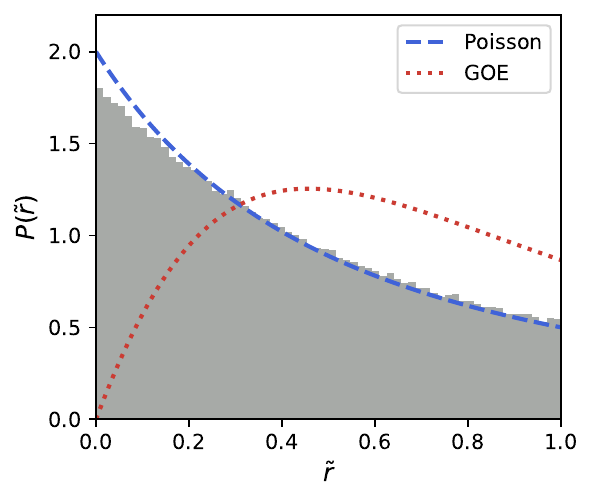}
\subfigimg[width=130pt]{\raisebox{0pt}{\quad\;\;\sublabel{k}}}{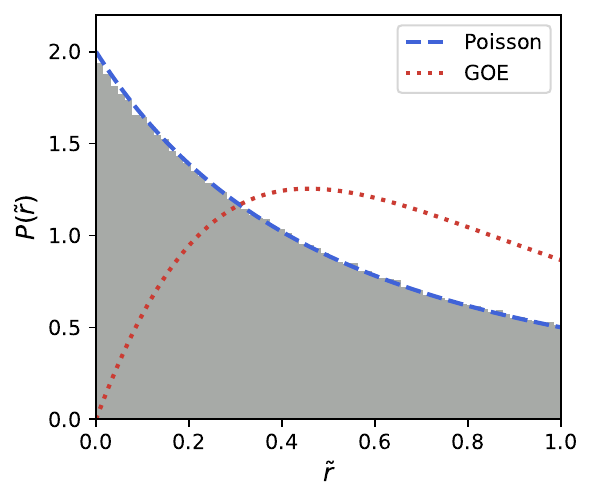}
\caption{\label{fig:ice-level-statistics-full}%
Level statistics for disordered $H_{\bowtie}$ on a 30-site lattice of shape $(3,-1)\times(1,3)$.
The panels show the probability density distributions $P(\tilde{r})$, for
\refsublabel{a}--\refsublabel{d} $\mathbb{Z}_2$ symmetric sector of connected components 1, 2, 3, and 4, respectively,
\refsublabel{e}--\refsublabel{h} $\mathbb{Z}_2$ antisymmetric sector of connected components 1, 2, 3, and 4, respectively,
\refsublabel{i} $\mathbb{Z}_2$ symmetric sector (combining all four connected components),
\refsublabel{j} $\mathbb{Z}_2$ antisymmetric sector,
and \refsublabel{k} the complete ice manifold.
The blue and red dashed curves respectively mark the $P(\tilde{r})$ for Poisson and GOE distributions.
For the GOE level statistics, we use the expression for $P(\tilde{r})$ from Ref.~\cite{atas_2013_prl}.
}
\end{figure*}

\para{}
In the main text we have presented the level statistics of the ice manifold in a 30-site lattice, within the $\mathbb{Z}_2$ symmetric sector (combining all four topological sectors) as well as within a single topological sector, also in the $\mathbb{Z}_2$ symmetric sector.
In fact, all sectors (four topological $\times$ two $\mathbb{Z}_2$) individually show almost identical GOE-like level statistics, shown in Figs.~\ref{fig:ice-level-statistics-full}(a)--\ref{fig:ice-level-statistics-full}(h).
Combining multiple sectors results in the level statistics appearing more Poisson-like [see Figs.~\ref{fig:ice-level-statistics-full}(i)--\ref{fig:ice-level-statistics-full}(k)]:
the larger the number of combined sectors, the closer the level statistics is to a Poisson distribution.
[Compare Figs.~\ref{fig:ice-level-statistics-full}(i), \ref{fig:ice-level-statistics-full}(j) with \ref{fig:ice-level-statistics-full}(k).]

\section{Stability of the ice gap of the Balents-Fisher-Girvin model}
\label{app:bfg-ice-gap}

\begin{figure}
\subfigimg[height=125pt]{\!\!\!\!\!\raisebox{4pt}{\sublabel{a}}}{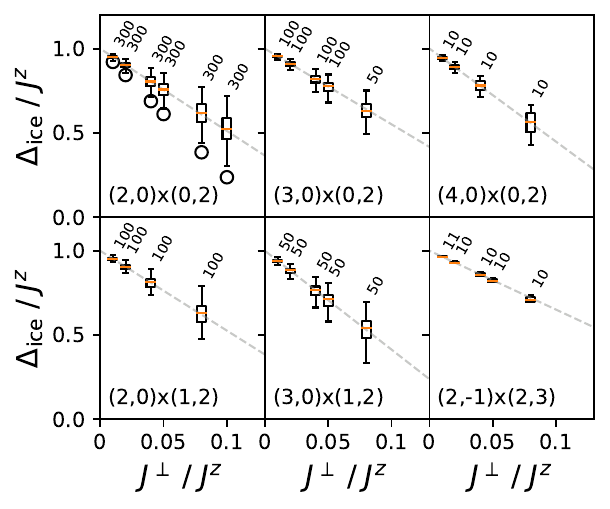}
\subfigimg[height=125pt]{\!\!\!\raisebox{4pt}{\sublabel{b}}}{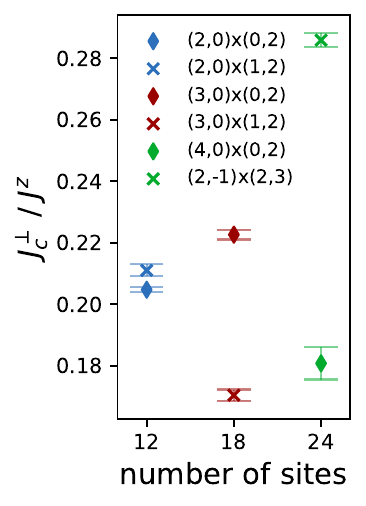}
\caption{\label{fig:ice-gap}
\refsublabel{a}
Box plots for the ``ice gap'' $\Delta_{\text{ice}}$, sampled over disorder configurations $\{J_{ij}^{\perp}\}$.
The box ranges from the lower to upper quartiles, with a line marking the median, and the whiskers showing the range of the data.
The number above a box indicates the number of sampled disorder configurations, and the dashed lines are fits to the disorder averaged $\Delta_{\mathrm{ice}}$ with the form
$\Delta_{\mathrm{ice}} /J^z = 1 - J^{\perp} / J^{\perp}_c$.
\refsublabel{b}
$J^{\perp}_c$ vs number of sites.
The horizontal lines mark the standard error for $J^{\perp}_c$.
}
\end{figure}

\para{}
The level statistics of the random-bond $H_{\text{BFG}}$ for $|J^{\perp}| \ll J^z$ on the whole ice manifold (even after resolving the $\bbZ_2$ spin-flip symmetry required by the ice rule) is expected to be Poisson-like, especially so in the thermodynamic limit since the level repulsions between different topological sectors are exponentially suppressed.
With increasing $|J^{\perp}|$, however, the ice gap $\Delta_{\mathrm{ice}}$---the energy separation between the ice manifold and the rest of spectrum---closes at $J^{\perp} = J_c^{\perp}$.
The level statistics crossovers from Poisson-like to GOE-like at scale $J^{\perp} / J_c^{\perp} \sim O(1)$.

\para{}
For the system sizes that we have considered,
the ice gap $\Delta_{\text{ice}}$ depends linearly on $J^{\perp}$ to leading order
[shown in Fig.~\ref{fig:ice-gap}(a)].
We can extract estimates for $J_c^{\perp}$ from linear fits to the means of $\Delta_{\mathrm{ice}}$ over different random configurations.
As shown in Fig.~\ref{fig:ice-gap}(b), we do not find a clear sign that $J_{c}^{\perp}$ vanishes with increasing system size.
If indeed the $J_{c}^{\perp}$ remains finite in the thermodynamic limit,
the level statistics is expected to show either a crossover or possibly a sharp transition between Poisson-like and GOE-like distributions, at a nonzero value of $J^{\perp}$.

\section{Kempe connectivity of three-coloring manifold}
\label{app:kempe-connectivity}

\begin{table}[b]
\caption{\label{tab:kempe-connectivity-81}%
Kempe connectivity of three-colorings on $(3,-3)\times(3,6)$ Kagome lattice (45184 colorings total).
}
\begin{tabularx}{\columnwidth}{cX}
\hline\hline
loop length & component size (number of components)\\
\hline
6  & 19298(1), 378(18), 180(3), 20(162), 19(108), 17(324),
     14(108), 13(162), 8(18), 4(162), 3(324), 2(486), 1(1388) \\
10 & 22052(1), 3942(2), 2214(6), 180(3), 3(324), 1(452)\\
12 & 37172(1), 3942(2), 1(128)\\
14 & 37172(1), 3942(2), 54(2), 1(20)\\
18 & 37172(1), 3942(2), 54(2), 20(1)\\
20 & 37280(1), 3942(2), 20(1)\\
22 & 37280(1), 7884(1), 20(1)\\
\hline\hline
\end{tabularx}
\end{table}

\begin{figure*}
\subfigimg[width=200pt]{\!\!\!\!\!\!\!\!\!\!\!\!\!\!\!\!\sublabel{a}}{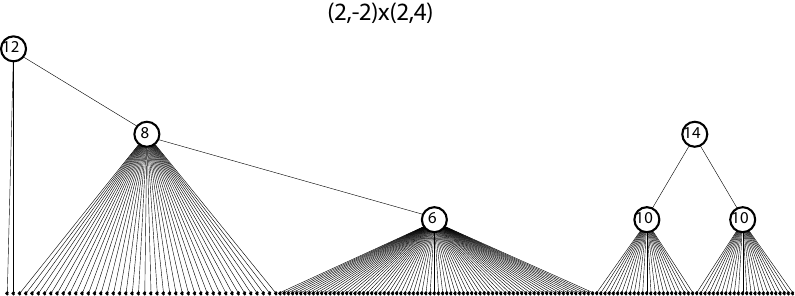}\\\vspace{0.5em}%
\subfigimg[width=500pt]{\!\!\!\!\!\!\!\!\!\!\sublabel{b}}{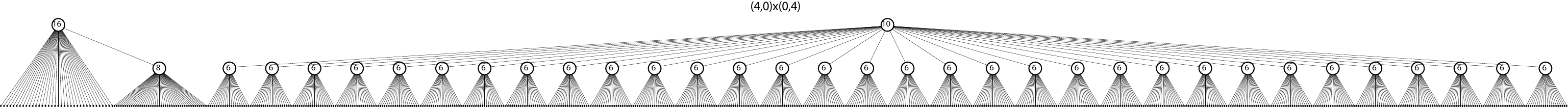}
\caption{\label{fig:kempe-hierarchy}%
Hierarchical connectivity structures of the three-coloring manifold with Kempe moves at various loop lengths, for
\refsublabel{a} 36-site kagome lattice of shape $(2,-2)\times(2,4)$, and
\refsublabel{b} 48-site lattice of shape $(4,0)\times(0,4)$.
The leaf nodes at the bottom are three-coloring configurations,
and the internal nodes labeled $\ell$ are connected components when loops of length $\ell$ or less are allowed, with increasing value of $\ell$ from bottom to top.
}
\end{figure*}

\begin{figure}\begin{minipage}{202pt}\flushright
\subfigimg[width=200pt]{\raisebox{-2pt}{\qquad\;\;\sublabel{a}}}{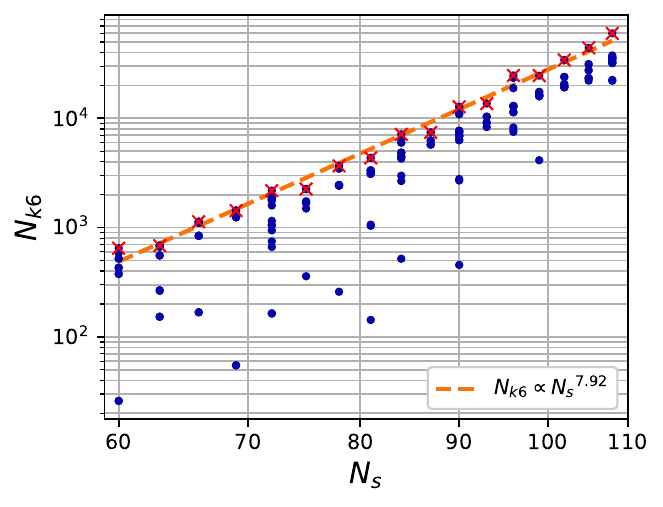}\qquad\\%
\subfigimg[width=192pt]{\raisebox{-2pt}{\quad\;\;\;\sublabel{b}}}{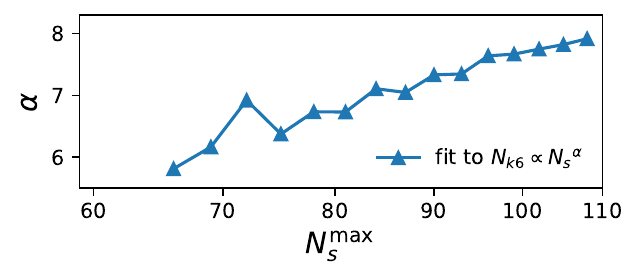}\qquad\end{minipage}
\caption{\label{fig:kempe-component-vs-size-power}%
\refsublabel{a}
Number of connected components in Kempe connectivity graph with loops of length 6 ($N_{k6}$) vs. number of sites ($N_s$), for various system geometries, plotted on a log-log scale.
The orange dashed line is a power-law fit $N_{k6} \sim {N_{s}}^{\alpha}$ to the largest number of components for every system size (marked by a red cross).
\refsublabel{b}
Scaling analyses for the power-law fits.
The $y$ axes represent parameters from fits to data points in the range $60 \le N_{s} \le N_{s}^{\mathrm{max}}$.
}
\end{figure}

\para{}
The connectivity structure of the three-coloring manifold changes with the lengths of the Kempe moves.
Table~\ref{tab:kempe-connectivity-81} shows the sizes of the connected components in an 81-site $(3,-3)\times(3,6)$ lattice, when Kempe loops of certain length or shorter are allowed.
At loop length 6, which is the shortest local loop length possible, the coloring manifold is still fragmented into 3264 sectors;
when loops of all lengths are allowed, these eventually coalesce into three large connected components.

\para{}
The hierarchical connectivity structure that arises from the length dependence are shown in
Fig.~\ref{fig:kempe-hierarchy} for 36-site and 48-site lattices.
The result for 81 sites is too large to plot.

\para{}
In the main text (Fig.~\ref{fig:three-coloring}) we showed that the number of fragments,
each consisting of three-coloring states connected by Kempe loop lengths of size six ($N_{k6}$), is exponential in system size $N_s$.
In Fig.~\ref{fig:kempe-component-vs-size-power}, we replot the same numerical data, but instead attempt to fit it to a polynomial form $N_{k6} \sim N_s^{\alpha}$. We find that
$\alpha$ changes with $N_s$, strongly suggesting that this functional form is incorrect.

\section{Kempe dynamics of three-coloring states}
\label{app:kempe-dynamics}

\para{}
At short times, the Kempe relaxation dynamics of a coloring state is fully determined by the number of other coloring states connected to it through Kempe moves.
Here we show that the Loschmidt echo approximates to
\begin{align}
  \left\vert \bracket{\psi(0), \psi(t)} \right\vert^2
  &
  \approx 1 - d \kappa_6^2 t^2
\end{align}
where $d$ is the number of other states connected to the initial coloring state, i.e. the degree in the Kempe connectivity graph.

\begin{figure*}
\subfigure[\label{fig:kempe-loschmidt-loglog}]{\includegraphics[height=100pt]{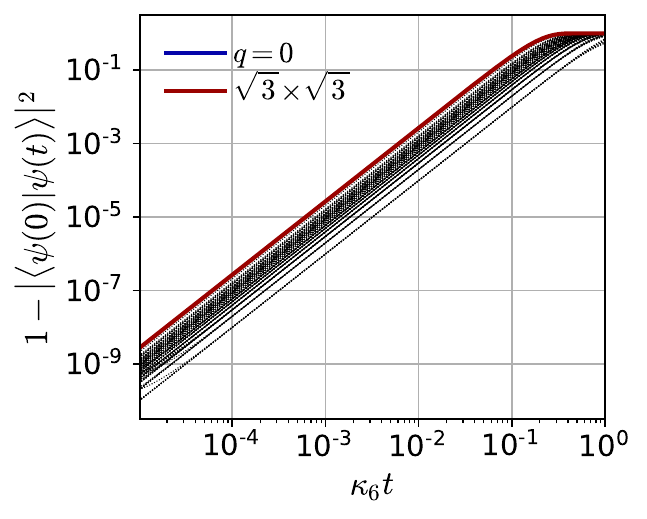}}
\subfigure[\label{fig:kempe-histogram-81}]{\includegraphics[height=107pt]{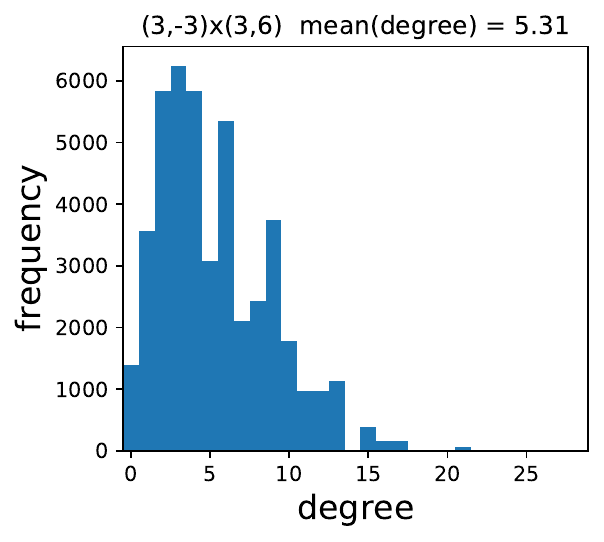}}%
\subfigure[\label{fig:kempe-histogram-108}]{\includegraphics[height=107pt]{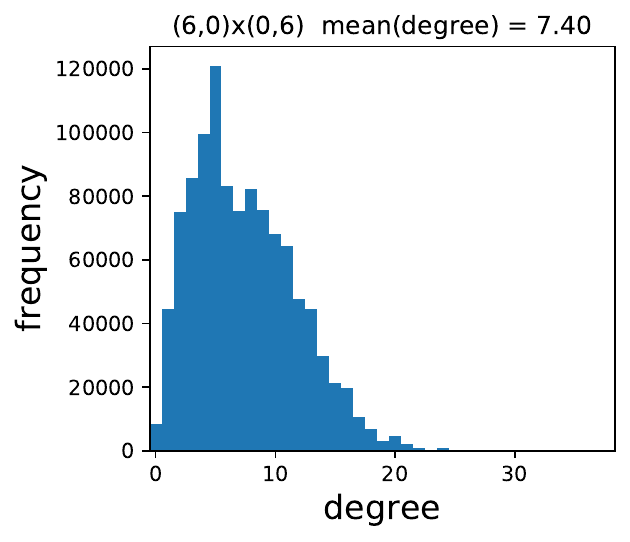}}%
\subfigure[\label{fig:kempe-histogram-degree-vs-size}]{\includegraphics[height=107pt]{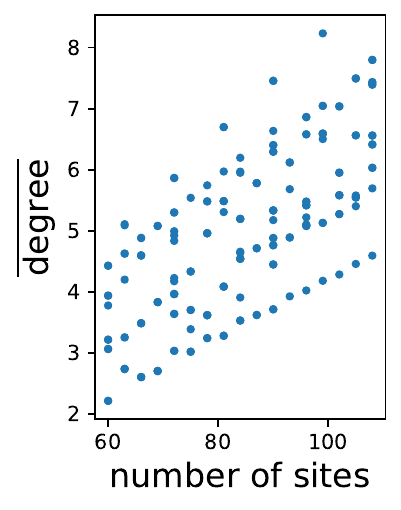}}
\caption{\label{fig:kempe-histogram}%
\refsublabel{a}
Loschmidt echoes for the length-6 Kempe model on an 81-site kagome lattice [same the data as Fig.~3(d)] plotted as $1 - \vert \langle \psi(0) \vert \psi(t) \rangle \vert^2$ vs. $\kappa_6 t$ in log-log scale.
For small $t$ ($\kappa_6 t \lesssim 0.1$), the curves follow $1 - \vert \langle \psi(0) \vert \psi(t) \rangle \vert^2 \propto t^2$.
\refsublabel{b},\refsublabel{c} Degree histograms of the Kempe connectivity graphs with loops of length 6 for
systems of shapes $(3,-3)\times(3,6)$ (81 sites) in \refsublabel{b} and  $(6,0)\times(0,6)$ (108 sites) in \refsublabel{c}.
\refsublabel{d} Average degree vs number of sites for various lattice geometries.
}
\end{figure*}

\para{}
Denote the initial coloring state as $\ket{\psi(0)} = \ket{C_{0}}$.
The Kempe Hamiltonian can then be broken into terms that act on $\ket{C_{0}}$, and those that do not:
\begin{align}
  H_{K}
    &=
      \kappa_{6} \sqrt{d}
        \left( \frac{1}{\sqrt{d}} \sum_{k \in K_{6}(C_0)}  \ket{k(C_0)} \right)
        \bra{C_0} \nonumber\\
    &\quad+ \kappa_{6} \sum_{C \neq C_0} \sum_{k \in K_{6}(C)} \ket{k(C)} \bra{C},
\end{align}
where $K_{6}(C)$ is the set of length-6 Kempe moves for the coloring $C$,
and $d \equiv \left\vert K_{6}(C_0) \right\vert$ is the number of moves for $C_0$.
At small $t$, the leading order time-evolution can be approximated by that of a two-level system,
whose states are $\ket{C_0}$ and $\ket{C_0'} \equiv \sum_{k}  \ket{k(C_0)} / \sqrt{d}$:
\begin{align}
  e^{\left(-i H_K t \right)} \ket{C_0}
    &\approx
      \cos \left( \sqrt{d} \kappa_6 t \right) \ket{C_0}
      - i \sin \left( \sqrt{d} \kappa_6 t \right) \ket{C_0'}
\end{align}
The Loschmidt echo therefore approximates to
\begin{align}
   \left\vert \bracket{\psi(0), \psi(t)} \right\vert^2
      &\approx \cos^2 \left( \sqrt{d} \kappa_6 t \right)
   \approx 1 - d \kappa_6^2 t^2.
\end{align}
The $t^2$ dependence is in agreement with the log-log plot in Fig.~\ref{fig:kempe-loschmidt-loglog}.

\para{}
The relationship between the number of Kempe moves and the dynamics allows us to study the statistics of relaxation time scale of the coloring states through graph analyses of Kempe connectivity.
Figures~\ref{fig:kempe-histogram}(b) and \ref{fig:kempe-histogram}(c) show degree histograms of Kempe connectivity graphs of loop length 6.
As Fig.~\ref{fig:kempe-histogram}(d) shows, the average degree appears to grow linearly with the system size.
This is in agreement with the fact that the maximum number of (two color) Kempe loops is that of the $\sqrt{3}\mathord{\times}\sqrt{3}$ state, and is given by the number of hexagonal plaquettes.%

\section{Dynamics of unprojected coloring states}
\label{app:unprojected-coloring}

\begin{figure}\centering
\includegraphics[width=160pt]{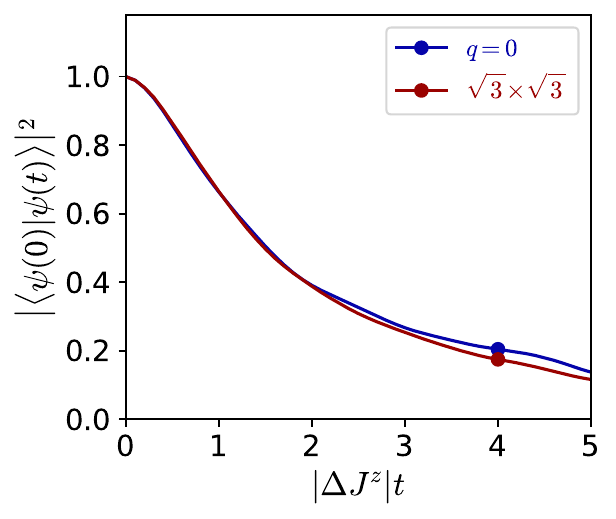}
\caption{\label{fig:loschmidt-xxz-18-unprojected}%
Loschmidt echo plot for $q\mathord{=}0$ and $\sqrt{3}\mathord{\times}\sqrt{3}$ states on
the 18-site lattice of shape $(3,0)\times(1,2)$, at $\Delta J^{z} \equiv J^{z} + 1/2 = 0.01$.
}
\end{figure}

\para{}
As pointed out in the main text, the three-coloring states, whether projected or unprojected, are eigenstates of the nearest-neighbor XXZ Hamiltonian at $J^z=-1/2$.
The unprojected coloring state is written as
\begin{align}
  \ket{\boldsymbol{\gamma}}
    \equiv
    \bigotimes_{i=1}^{N} \ket{\gamma_{i}},
\text{ where }
  \ket{\gamma_{i}}
    &=
      \frac{1}{\sqrt{2}}
      \left(
        \ket{\up} + e^{i\alpha_{i}} \ket{\dn}
      \right)
\end{align}
with $\alpha_{i} = 0, \frac{2\pi}{3}, \frac{4\pi}{3}$ for the three colors.
The projected states can be written as
\begin{align}
  \ket{\boldsymbol{\gamma}, m} &\equiv \frac{1}{\calN_{m}} P_{S_z=m} \ket{\boldsymbol{\gamma}}.
\end{align}
These coloring states, however, are not mutually orthogonal.
Since
\begin{align}
  \bracket{\gamma_{i}', \gamma_{i}}
    =
      \frac{1 + e^{i (\alpha_{i} - \alpha_{i}') }}{2}
    =
      1, \frac{e^{\pm \frac{i 2 \pi}{3}}}{2},
\end{align}
the overlap between two colorings states is written as
\begin{align}
  \left\vert \bracket{ \boldsymbol{\gamma}', \boldsymbol{\gamma} } \right\vert
    &=
      \frac{1}{2^{n_{d}}}
  \label{eq:overlap-unprojected}
\end{align}
where $n_{d}$ is the number of sites with different colors between $\boldsymbol{\gamma}$ and $\boldsymbol{\gamma}'$.
In other words, two coloring states can only be orthogonal to each other in the thermodynamic limit.

\para{}
Overlap between projected coloring states, on the other hand, is sector dependent.
In the extreme limit of fully polarized sectors, all coloring states project to the exact same state $\ket{\up\up\cdots}$ or $\ket{\dn\dn\cdots}$.
In the unpolarized sector (total $S_z=0$ or $\pm 1/2$), on the other hand, the overlaps between different colorings are less than unity,
and can even be smaller than the overlap between unprojected states in Eq.~\eqref{eq:overlap-unprojected}.

\para{}
At points away from $J^z=-1/2$, these states are no longer eigenstates of the Hamiltonian, and thus evolve with time.
How do these nonzero overlaps (i.e., the nonorthogonal nature of the coloring manifold) affect the dynamics?
Would the unprojected coloring states relax faster compared to the projected states since there are more overlaps between them, which translates to more decay channels?
Would having more channels reduce the difference between different coloring states?

\para{}
We show the Loschmidt echo $|\bracket{\psi(0), \psi(t)}|^2$ for the unprojected $\ket{q\mathord{=}0}$ and $\ket{\sqrt{3}\mathord{\times}\sqrt{3}}$ states in Fig.~\ref{fig:loschmidt-xxz-18-unprojected}.
The two states show almost identical relaxation.
This is in contrast to the projected coloring states $P_{S_z=m} \ket{q\mathord{=}0}$ and $P_{S_z=m}\ket{\sqrt{3}\mathord{\times}\sqrt{3}}$ presented in the main text, where there is a stark difference between the two states.
Also, the relaxation timescale is shorter compared to that of the projected states by a factor of $\sim 5$.

\section{Constructing Kempe connectivity graphs}
\label{app:kempe-algorithm}

\para{}
Here we describe exactly how we construct the Kempe connectivity graph for the three-coloring manifold on kagome lattices.
As we clarified in the paper, we define the colorings as ``relative colorings,'' meaning that two colorings are equivalent if they differ by a global color rotation, analogous to projecting to a particular $S_z$ sector.
To enforce this equivalence numerically, we define a ``normal'' coloring by fixing the color of a particular site.
More precisely, given a three-coloring $c: V \rightarrow \mathbb{Z}_3$ where $V$ is the set of sites, the normalized coloring $c'$ is defined as $c': v \mapsto c(v) - c(v_0) \text{ mod } 3$,
for an arbitrarily (but consistently) chosen $v_0 \in V$.

\para{}
For any three-coloring of a kagome lattice, every nearest-neighbor bond belongs to a single Kempe loop.
This property is specific to kagome lattice, where every node has degree 4.
(In the triangular lattice, for example, a bond belongs to multiple Kempe loops.)
We can make use of this nonbranching property to find all Kempe loops of all three-colorings.
Given a coloring and a bond, Algorithm~\ref{alg:follow-kempe-loop} finds the Kempe loop that the bond belongs to.
Algorithm~\ref{alg:find-kempe-loops} can then be used to find all Kempe loops of a given coloring.

\para{}
Using these, we can construct the Kempe connectivity graph, whose vertices are the three-colorings of the lattice,  with an edge between two colorings if there is a Kempe move that connects the two.
Algorithm~\ref{alg:kempe-connectivity-graph} describes the steps.

\begin{algorithm}
\begin{singlespace}
\DontPrintSemicolon
  \KwData{ $G = (V,E)$, undirected graph representing the lattice,\\
\qquad\quad$c: V \rightarrow \mathbb{Z}_{3}$, a three-coloring of $G$,\\
\qquad\quad$(u,v) \in E$
}
  \KwResult{$\ell$, a list of sites in the Kempe loop that includes $(u,v)$}
  \Begin{
    $\ell \gets (u, v)$\;
    $(\tilde{c}_1, \tilde{c}_2) \gets (c(u), c(v))$\;
    $w \gets v$ \;
    \While{$w \neq u$}{
      $w \gets$ a neighbor of $w$ with color $c(x) = \tilde{c}_{1}$ that is not in $\ell$\;
      append $w$ to $\ell$\;
      swap $\tilde{c}_1$ and $\tilde{c}_2$\;
    }
  }
\end{singlespace}
\caption{\textsc{FollowKempeLoop}$(G, c, (u,v))$}
\label{alg:follow-kempe-loop}
\end{algorithm}

\begin{algorithm}
\begin{singlespace}
\DontPrintSemicolon
  \KwData{$G=(V,E)$, undirected graph representing the lattice\\
\qquad\quad$c: V \rightarrow \mathbb{Z}_{3}$, a three-coloring of $G$
  }
  \KwResult{$L$, set of all Kempe loops of $c$}
  \Begin{
    $S \gets E$\;
    $L \gets \emptyset$\;
    \While{$S$ is not empty}{
      $b \gets$ an element of $S$\;
      $\ell \gets$ \textsc{FollowKempeLoop}($G, c, b$)\;
      remove all bonds of $\ell$ from $S$\;
      add $\ell$ to $L$\;
    }
  }
\end{singlespace}
\caption{\textsc{FindKempeLoops}$(G, c)$}
\label{alg:find-kempe-loops}
\end{algorithm}

\begin{algorithm}
\begin{singlespace}
\DontPrintSemicolon
  \KwData{$G=(V,E)$, undirected graph representing the lattice}
  \KwResult{$G_K = (V_K, E_K)$, the Kempe connectivity graph}
  \Begin{
    $C \gets$ the set of all $n$-colorings of $G$\;
    \ForEach{$c \in C$}{
      $L \gets$ \textsc{FindKempeLoops}$(G, c)$\;
      \ForEach{$\ell \in L$}{
        $c' \gets$ flip the colors of vertices in $\ell$ from $c$\;
        $c'' \gets$ normalize coloring $c'$\;
        Add $(c,c'')$ to $E_K$
      }
    }
  }
\end{singlespace}
\caption{\textsc{KempeConnectivityGraph}$(G)$}
\label{alg:kempe-connectivity-graph}
\end{algorithm}

\bibliography{refs}

\end{document}